\documentclass[%
preprint,
 amsmath,amssymb,
floatfix,
]{revtex4-2}

\usepackage{graphicx}
\usepackage{dcolumn}
\usepackage{bm}
\usepackage{hyperref}
\newcommand{\sq}[1]{\left[ #1 \right] }
\newcommand{\rnd}[1]{\left( #1 \right) }
\newcommand{\bra}[1]{\left\{ #1 \right\}}
\newcommand{\abs}[1]{\left\vert #1 \right\vert}

\renewcommand{\vec}[1]{\mathbf{#1}} 
\usepackage{times}
\usepackage{textcomp}
\usepackage[ruled,linesnumbered]{algorithm2e}

\usepackage{placeins}
\usepackage{nicefrac}
\usepackage[super]{nth}
\usepackage{gensymb}
\usepackage{color}
\usepackage{xcolor}
\newcommand{\red}[ 1][]{\textcolor{red}}
\definecolor{myblue}{rgb}{0.0, 0.375, 0.75}  
\newcommand{\blue}[ 1][]{\textcolor{myblue}}
\definecolor{mygreen}{rgb}{0.2, 0.55, 0}  
\newcommand{\green}[ 1][]{\textcolor{mygreen}}

\renewcommand{\red}[ 1][]{\textcolor{red}}
\definecolor{TPcol}{rgb}{0.847, 0.353, 0.102}  
\definecolor{AHcol}{rgb}{0.0, 0.375, 0.75}  
\definecolor{JHcol}{rgb}{0.2, 0.55, 0}  
\definecolor{AGcol}{rgb}{1, 0.4, 0.55}  
\newcommand{\AH}[ 1][]{\textcolor{AHcol}}
\newcommand{\TP}[ 1][]{\textcolor{TPcol}}
\newcommand{\JH}[ 1][]{\textcolor{JHcol}}
\newcommand{\AG}[1][]{\textcolor{AGcol}}

\begin{document}


\title{Prediction of thermal conductivity in dielectrics\\ using fast, spectrally-resolved phonon transport simulations}

\author{Jackson R. Harter}
\email{harterj@oregonstate.edu}
\homepage{https://rtrp.github.io/osu-transport/users/harterj/}
\affiliation{Oregon State University, Corvallis, OR 97330}
\affiliation{Idaho National Laboratory, Idaho Falls, ID 83402}
\author{S. Aria Hosseini}%
\affiliation{University of California - Riverside, Riverside, CA 92521}
\author{Todd S. Palmer}
\email{palmerts@engr.orst.edu}
\homepage{https://rtrp.github.io/osu-transport/}
\affiliation{School of Nuclear Science and Engineering \\ Oregon State University, Corvallis, OR 97330}
\author{P. Alex Greaney}
\email{greaney@ucr.edu}
\homepage{http://alexgreaney.com}
\affiliation{Department of Mechanical Engineering\\ University of California - Riverside, Riverside, CA 92521}

\date{\today}

\begin{abstract}
We present a new method for predicting effective thermal conductivity ($\kappa_{\textrm{eff}}$) in materials, informed by \emph{ab initio} material property simulations. Using the Boltzmann transport equation in a Self-Adjoint Angular Flux formulation, we performed simulations in silicon at room temperatures over length scales varying from 10 nm to 10 $\mu$m and report temperature distributions, spectral heat flux and thermal conductivity. Our implementation utilizes a Richardson iteration on a modified version of the phonon scattering source. In this method, a closure term is introduced to the transport equation which acts as a redistribution kernel for the total energy bath of the system. This term is an effective indicator of the degree of disorder between the spectral phonon radiance and the angular phonon intensity of the transport system. We employ polarization, density of states and full dispersion spectra to resolve thermal conductivity with numerous angular and spatial discretizations.
\end{abstract}

\maketitle

\section{\label{sec:intro}Introduction}
In many applications a material's propensity for conducting heat is an important factor in its overall performance. For thermal management applications, having high or low thermal conductivity is of principal importance, but thermal conductivity makes an auxiliary contribution to a material's figure of merit in many other applications such as batteries, high frequency electronics, and thermoelectric materials.

In order to accelerate the discovery, design, and selection of application-driven materials, it is desirable to efficiently compute a thermal conductivity from first principles. The past decade has seen considerable progress towards this goal in semiconductors and ceramics, where heat transport is dominated by phonons~\cite{phononMinnich,FengRuanReview}. Methods have been developed which make routine the computing of second order interatomic stiffness matrices of a crystal from density functional theory (DFT) calculations, and from them, the phonon dispersion~\cite{vasp1,vasp2,Carrete_2017,ShengBTE_2014,phonopy,phono3py}. More recently, methods have been developed for computing third and fourth order stiffness matrices from which the rates of intrinsic three- and four-phonon scattering processes may be derived. The use of these methods is becoming commonplace, as together they enable one to compute the intrinsic phonon thermal conductivity, $\kappa$, using the equation derived from the Boltzmann transport equation for a phonon gas:
\begin{equation} 
\label{eq:kappa-BZint}
     \kappa = \frac{1}{8\pi^3} \sum_p \int_{\mathrm{BZ}} \tau_{p\vec{k}} \vec{v}_{p\vec{k}} \otimes \vec{v}_{p\vec{k}} \omega_{p\vec{k}} \hbar n\left(\omega_{p\vec{k}},T \right) d\vec{k}^3,
\end{equation}
which integrates the contributions to thermal transport from all phonons at all polarizations $p$ and wave vectors $\vec{k}$ in the crystal's first Brillouin zone. Here, $\omega$ and $\vec{v}_{p,\vec{k}}$ are the phonon angular frequency and group velocity, respectively, which are obtained from the phonon dispersion relation. 

The equation above gives a material's inherent capacity for transporting heat under the assumption that the perturbation in the phonon population that gives rise to a net heat flux is spatially homogeneous. However, in most practical applications materials contain defects and impurities in addition to micro and nano-structured morphology; these all present additional resistance to phonon transport. Predicting the macroscale thermal conductivity in these heterogeneous materials is more complicated, but is of profound technological importance as the hierarchy of defect structure and morphology can be central to the material's functionality. For example, in thermoelectric devices, dopants, impurities, inclusions, and nano-structuring are engineered into materials in order to optimize the electrical and thermal properties. In other applications where materials are under irradiation, it is essential to predict the evolution of thermal conductivity during the accumulation of radiation-induced damage. 

In materials where atomistic to micro-scale heterogeneity impacts thermal conductivity, the phonon population is also not spatially uniform, and so predicting the macroscopic thermal conductivity requires an understanding of the full phonon population distribution within some representative microstructure. This can be done by \emph{simulating} the spectrally-dependent phonon distribution as governed by the Boltzmann transport equation (BTE) using phonon transport properties derived from DFT calculations of atomic interactions. The BTE describes the evolution of a distribution of particles as a function of their position, frequency and direction of travel, making the BTE a six-dimensional equation (or seven-dimensional, in the time-dependent case). 

Advances have been made in the simulation of spectrally-dependent phonon transport, providing important insights into the spectral contribution to heat transport, but to date the computational burden associated with BTE solutions have limited the size, dimensionality, and/or physics of the systems considered. Allu \emph{et al.}~\cite{mfAllu1} investigated frequency dependence of the BTE in silicon slabs and reported spectral thermal conductivity based on Knudsen number (the mean free path divided by the domain length). However, the trade-off of this approach was their reported high computational expense, and the discrete phonon groups had no coupling. Mazumder and Majumdar~\cite{mcMazumder} demonstrated polarization dependence of phonons in Monte Carlo transport simulations and showed the importance of the branch dependence; polarization transitions occur for background scattering as well as anharmonic interactions~\cite{anharmSharma}. Hua derived the time and frequency dependent phonon transport equation using the method of degenerate kernels in an attempt to increase computational efficiency compared to traditional integral discretization methods~\cite{mfHua}. This was only done in 1-D, did not provide any spatial discretization, and was primarily analytical. As simulating frequency dependent phonon transport allows for the characterization of the frequency spectra, it allows for mode dependent contributions to the specific heat capacity and phonon velocity. Minnich~\emph{et al.}~\cite{mfMinnich} demonstrated this mode dependence in their simulations of aluminum and silicon thin films, and calculated the spectral heat capacity from each frequency interval, explicitly illustrating the contribution of each mode to thermal conductivity. Rather than discriminating phonons by frequency and polarization, Romano has developed an approach that separates phonon modes into groups by discretizing the spectrum of phonon mean free path (MFP)~\cite{Romano1,Romano2}. This provides considerable computational savings as the MFP is the only spectrally relevant phonon property that appears in the BTE. Most recently, Zhang \emph{et al.}~\cite{zhangBTE} developed a coupled approach using a reference temperature formulation which efficiently simulates temperature distribution and thermal conductivity over the nano- to micro-scale length range. However, they did not consider optical phonons with short mean free paths in their simulations. We discuss the results of Zhang~\emph{et al.} in greater detail below, but taken together, these prior works highlight that for simulating phonons it is imperative that the phonons are spectrally-resolved, and that however one defines the phonon spectrum, one cannot cavalierly disregard portions of it. This makes fully self-coupled calculations of large systems computationally challenging. In this manuscript we present an alternative approach for solving the spectrally-dependent phonon BTE that draws on numerical methods developed for neutron transport in the software package Rattlesnake~\cite{rsm}. The method solves the BTE in the self-adjoint angular flux (SAAF) formulation, which is computationally efficient, making it possible to simulate phonon transport in relatively large systems.

The numerical procedures for simulating the collective transport of phonons occupying a spectrum of vibrational modes is similar to the procedures used in nuclear engineering for simulating multi-group neutron or radiation transport~\cite{DandH,radMorel}. With neutron or radiation transport, the energy spectrum is partitioned into discrete groups and the transport simulation accounts for within-group and group-to-group scattering of neutrons based on the interaction cross sections. Rattlesnake~\cite{rsm} employs a multi-group energy discretization scheme to solves the BTE for neutrons. It is developed in the Multi-physics Object Oriented Simulation Environment (MOOSE) framework~\cite{Gaston} at Idaho National Laboratory (INL) and solves the SAAF-BTE using a finite element spatial discretization and a discrete ordinates angular discretization. Having previously adapted Rattlesnake to model one-group (grey) phonon transport~\cite{HarterANS2015,HarterANS2016,HarterJHT2018} in both homogeneous and heterogeneous media, in the work presented here we have further developed Rattlesnake to simulate spectrally-dependent phonon transport.

The remainder of the manuscript is organized as follows: In the next section we develop the methods used to derive the SAAF-BTE for spectral transport, temperature coupling between discrete phonon groups, and the energy redistribution function used in closing the system of equations. Following this, we describe the methods used to compute a material's phonon properties and the methods we have developed to convolve these data into the collective properties of a set of discrete transport groups simulated in the SAAF-BTE. Finally, to demonstrate the computational efficiency of our multi-group, temperature coupled approach, simulations are presented of phonon transport in silicon of varying thickness. For these we report spectrally-resolved temperature, heat flux, and thermal conductivity. We discuss each of these quantities, how they are affected by spatial and angular resolution, the effect of the new closure term, and how convergence is affected by the simulation of all phonon modes.

\section{Methods}
The generalized Boltzmann transport equation is used widely by the transport community to model the evolution of a distribution of particles or carriers. The modes of atomic vibration of crystal are traveling waves with well defined wave vectors. These waves extend across the entirety of the crystal; however, if uncertainty is applied to the wavevector the vibrations become propagating localized wavepackets. As the energy of vibrations comes in discrete quanta proportional to their vibrational frequency, individual wavepackets can be treated as quasiparticles -- phonons -- that collectively follow Bose-Einstein statistics. The total vibrational energy of the crystal can thus be modeled as a phonon quasiparticle gas and so can be described by the Boltzmann transport equation for uncharged particles:
\begin{equation}
\frac{df\rnd{\vec{r},\vec{k},p}}{dt} =  -\vec{v}\rnd{\vec{k},p}\cdot \vec{\nabla}f\rnd{\vec{r},\vec{k},p} + \left[\frac{df}{dt}\right]_\textrm{coll.}.
\end{equation}
Here $f\rnd{\vec{r},\vec{k},p}$ is the number of phonons at location $\vec{r}$ occupying the phonon mode with wavevector $\vec{k}$ and polarization $p$. The term $\vec{v}$ is the group velocity of a phonon wavepacket and is given by $\vec{v}=\vec{v}\rnd{\vec{k},p} = \nabla_{\vec{k}}\omega\rnd{\vec{k}}$, the derivative of the vibration's angular frequency, $\omega$, with respect to $\vec{k}$. 
The group velocity is a vector normal to the isofrequency surface for the dispersion relation which is not necessarily parallel to $\vec{k}$.
While the natural parameter for enumerating the phonon modes of a crystal is the wavevector and polarization (4 quantum numbers are required), for transport simulations it is more natural to parameterize the particle/carrier population by the particles' direction of travel, and then by their speed, and to treat these two parameters separately. This is particularly true for neutron transport as their dispersion relation is isotropic. To put the phonon transport problem in to a form where it can leverage the numerical power of existing transport solvers such as Rattlesnake, and restricting our interest to the time independent case, in what follows we write the Boltzmann transport 
\begin{equation}
    \label{eq:phononBTE}
    \left|\vec{v}\rnd{\hat{\vec{\Omega}},\eta,p}\right|\hat{\vec{\Omega}}\cdot \vec{\nabla}f\rnd{r,\hat{\vec{\Omega}},\eta,p} = \left[\frac{df\rnd{r,\hat{\vec{\Omega}},\eta,p}}{dt}\right]_\textrm{coll.},
\end{equation}
where $f=f(\vec{r},\hat{\vec{\Omega}},\eta,p)$ has been parameterized by the phonons' direction of propagation, $\vec{\hat{\Omega}}$, and spectral variable $\eta$.  The spectral metric $\eta$ could be energy, frequency, wave number, mean free path, or any other descriptor or quantum number for enumerating the phonon modes. 

The term on the right hand side (RHS) of Eq.~(\ref{eq:phononBTE}) is the rate of change of the population of due to the creation/annihilation of phonons and events that change a phonon's direction of travel. We model this using the single mode relaxation time (SMRT) approximation~\cite{Ziman} which assumes that the net rate of loss/gain of phonons in a single mode is described by a single relaxation time, $\tau = \tau(\hat{\vec{\Omega}},\eta,p)$, and is proportional to the deviation of the population from thermodynamic equilibrium. Defining the mean free path (MFP) of phonons $\Lambda(\hat{\vec{\Omega}},\eta,p) = \left|\vec{v}\right| \tau$,
the simplified BTE is written as
\begin{equation}
\label{eq:phononBTE-RTA}
\hat{\vec{\Omega}}\cdot\vec{\nabla}f = \frac{f^{0}\rnd{T\rnd{\vec{r}},\hat{\vec{\Omega}},\eta,p} - f}{\Lambda},
\end{equation}
where the equilibrium phonon distribution $f^{0}$ is described by Bose-Einstein statistics
\begin{equation*}
f^{0}=f^{0}\rnd{T\rnd{\vec{r}},\hat{\vec{\Omega}},\eta,p} = \langle n_\textrm{BE}  \rangle = \frac{1}{\exp\left[ \frac{\hbar \omega\rnd{\hat{\vec{\Omega}},\eta,p}}{k_\text{B}T\left(\vec{r}\right)}  \right]-1}.
\end{equation*}
Here $T\left(\vec{r}\right)$ is the local temperature,  $\hbar$ is the reduced Planck constant and $k_{B}$ is Boltzmann's constant. 

The population $f$ gives the \emph{number} of phonons in the mode characterized by $\vec{\hat{\Omega}}$, $\eta$, and $p$ and is thus dimensionless. 
As we are interested transport of heat rather than phonon number is more useful to weight the population by the modes' contribution to energy flux, and solve for the \emph{angular flux intensity}, $\psi$, which has dimensions of power per area per steradian, and is defined as
\begin{equation}
    \label{eq:intensityDefinition0}
    \psi\rnd{\vec{r},\hat{\vec{\Omega}},\eta,p} = \hbar \omega\vec{v}f\mathbb{D}d\eta.
\end{equation}
Here the term, $\mathbb{D}= \mathbb{D}(\hat{\vec{\Omega}},\eta,p)$ is the phonon density of states (DOS) per increment of solid angle $\vec{\hat{\Omega}}$ and spectral descriptor. 
It has dimensions of number per unit volume per steradian per $\eta$, and it satisfies the relationship
\begin{equation}
\label{eq:energyBalance0}
3N_v =\sum_p \int d\eta \int_{4\pi} \mathbb{D}\rnd{\hat{\vec{\Omega}},\eta,p}d\Omega,
\end{equation}
where $N_v$ is the number of of atoms per unit volume. The BTE in terms of angular radiance is
\begin{equation}
\label{eq:phononBTE-psi}
\hat{\vec{\Omega}}\cdot \vec{\nabla}\psi = \frac{\phi^{0} - \psi}{\Lambda},
\end{equation}
where $\phi^{0}$ is the \emph{radiance} of the mode at equilibrium given by
\begin{equation}\label{eq:radianceDefinition}
\phi^{0} = \phi^{0}\rnd{T\rnd{\vec{r}},\hat{\vec{\Omega}},\eta,p}
 = \frac{\hbar \omega~\vec{v}~ \mathbb{D}d\eta}{\exp\left[\frac{\hbar \omega}{k_\text{B}T\rnd{\vec{r}}}\right]-1},
\end{equation}
and is discretized into partitions with dependency on $\hat{\vec{\Omega}}$ within the Brillouin zone; this method is discussed in a proceeding section. The local temperature is defined as the temperature that a system at equilibrium would have in order to have the same thermal energy, $U_T$, as the phonon bath
\begin{eqnarray}
\label{eq:thermalenergy}
U_T\rnd{\vec{r}} &=& \sum_p\int d\eta \int_{4\pi} d\vec{\hat{\Omega}} \frac{\hbar \omega~ f~ \mathbb{D}}{\exp\left[\frac{\hbar \omega}{k_\text{B}T\rnd{\vec{r}}}\right]-1}\nonumber\\
&=& \sum_p \int d\eta \int_{4\pi} d\vec{\hat{\Omega}} \frac{\phi^{0}}{\left|\vec{v}\right|}.
\end{eqnarray}
The numerical approach to extract $T(\vec{r})$ from this relationship is described in detail later in the manuscript, but we first address the issue of energy conservation. 

The SMRT approximation for the collision term that appears in the BTE in Eqs.~(\ref{eq:phononBTE-RTA}) and (\ref{eq:phononBTE-psi}) assumes that modes which are out of equilibrium exchange energy with an implied phonon reservoir at the local temperature to relax them back into equilibrium. However, as the rate of this energy exchange is different for all modes for any arbitrary phonon population, it is likely that there will be a net flow of energy to or from the phonon bath, resulting in a lack of local energy conservation. The problem is that the flow of energy into the implied phonon bath in the SMRT is not reflected in the explicit phonon bath used to compute temperature in Eq.~(\ref{eq:thermalenergy}), and so we must include an additional source term to correct this.
\\
The net energy flow per unit volume, $\dot{U}_I$, to the implied phonon bath is
\begin{equation}
\label{eq:energyBalance1}
\dot{U}_I\rnd{\vec{r}} = \sum_p \int d\eta \int_{4\pi} d\Omega ~ \frac{\phi^0-\psi}{\Lambda}.
\end{equation}
To conserve energy, we must return this energy to the explicit phonon bath, and to do that we must first decide how to distribute the energy over the phonon modes. The most obvious choice for this is to add the energy back proportional to the equilibrium distribution of energy across the modes.
Applying this correction to the BTE to close the coupling of SMRT with the phonon bath gives:
\begin{equation*}
\label{eq:phononBTE-psi-closed}
\hat{\vec{\Omega}}\cdot \vec{\nabla}\psi = \frac{\phi^{0}- \psi}{\Lambda}-\beta\frac{\phi^0}{\left|\vec{v}\right|},
\end{equation*}
where $\beta$ is a rate and is given by
\begin{equation*}
\label{eq:beta}
\beta = \beta\rnd{\vec{r}}=\frac{\dot{U}_I}{U_T}. 
\end{equation*}
This residual energy projection is a measure of the balance of the equilibrium phonon population to the transport system population; $\beta$ is tightly coupled to temperature, the transport and equilibrium systems, and is the closure required to ensure energy conservation. In our framework, both the temperature and $\beta$ are functionals of the entire phonon population. The energy flow $\dot{U}_I$ can be either positive or negative depending on the phonon population. The most ballistic groups tend to have the strongest non-equilibrium condition occurring near isothermal boundaries and these are the regions in which we find $\beta$ to be the largest.

\subsection{SAAF-BTE for phonons}
In our previous work~\cite{HarterANS2015,HarterANS2016,HarterJHT2018}, we modified the SAAF-BTE for transport of grey neutrons to make it treat grey phonon transport, and then solved it to model heat conduction in UO$_2$ containing Xe bubbles. In grey transport the particles of interest are resolved in direction but the spectral properties are averaged over the entire distribution making a ``grey'' spectrum. In spectrally-resolved transport the grey approximation is equivalent to coupling the particle distribution evolution over angle, but having no coupling between the different portions of the property spectrum as in the first and second order equations
\begin{equation}
    \label{eq:oldRS-1}
    \hat{\vec{\Omega}}\cdot \nabla \psi\rnd{\vec{r},\hat{\vec{\Omega}}} + \frac{1}{\Lambda}\psi \rnd{\vec{r},\hat{\vec{\Omega}}} = \frac{1}{4\pi \Lambda} \phi^{\textrm{T}}\rnd{\vec{r}},
\end{equation}
\begin{equation}
    \label{eq:oldRS-2}
    -\hat{\vec{\Omega}}\cdot \nabla\sq{\Lambda\hat{\vec{\Omega}}\cdot \nabla \psi\rnd{\vec{r},\hat{\vec{\Omega}}} } + \frac{1}{\Lambda}\psi\rnd{\vec{r},\hat{\vec{\Omega}}} = \frac{1}{4\pi}\sq{\frac{1}{\Lambda}\phi^{\textrm{T}}\rnd{\vec{r}} - \hat{\vec{\Omega}}\cdot \nabla \phi^{\textrm{T}}\rnd{\vec{r}}}.
\end{equation}
Here the coupling term, $\phi^{\textrm{T}}$, is the zeroth angular moment
\begin{equation}
\label{eq:zeroMoment}
\phi^{\textrm{T}} = \phi^{\textrm{T}}\rnd{\vec{r}} = \int_{4\pi} \psi\rnd{\vec{r} ,\hat{\vec{\Omega}}} d\Omega,
\end{equation}
which we refer to as the \emph{transport scalar flux}. The total scalar flux is defined as the integral of the zeroth angular moment over all spectral groups and polarizations
\begin{equation*}
\label{eq:totalScalar}
    \Phi^\textrm{T} = \Phi^\textrm{T}\rnd{\vec{r}} =\sum_{p} \int_{\eta} \int_{4\pi}\psi\rnd{\vec{r} ,\hat{\vec{\Omega}}} d\Omega d\eta,
\end{equation*}
and $\phi^{\textrm{T}}$ and $\Phi^{\textrm{T}}$ both have units of $\textrm{W}\cdot\textrm{rad}^{-1}\cdot \textrm{m}^{-2}$.
In the BTE, the phonon angular intensity is defined along a direction of travel $\hat{\vec{\Omega}}$, a function of polar and azimuthal angles $\theta$ and $\phi$; $\hat{\vec{\Omega}}$ is an independent variable of $\psi$, which itself is the dependent variable, $\psi(\vec{r},\hat{\vec{\Omega}})$.

The weak form of Eq.~(\ref{eq:oldRS-2}) was then solved with Rattlesnake. In this current work, we introduce a new form of the SAAF-BTE specifically for phonons, which relies on implicit temperature coupling to better represent the subtle physics of phonon transport. We return to Eq.~(\ref{eq:phononBTE-RTA}), with phonon state variables multiplied through and the addition of the non-equilibrium bath source correction to the right hand side
\begin{equation*}
    \abs{\vec{v}} \hat{\vec{\Omega}} \cdot \nabla \psi\rnd{\vec{r},\hat{\vec{\Omega}}} = \frac{\frac{1}{4\pi}\phi^{0}\rnd{T\rnd{\vec{r}}}-\psi\rnd{\vec{r},\hat{\vec{\Omega}}}}{\tau} - \frac{\beta\rnd{T\rnd{\vec{r}}}\phi^{0}\rnd{T\rnd{\vec{r}}}}{4\pi},
\end{equation*}
and rearrange
\begin{equation}
\label{eq:firstOrderRearrange}
\hat{\vec{\Omega}}\cdot \nabla \psi\rnd{\vec{r},\hat{\vec{\Omega}}} + \frac{1}{\Lambda}\psi\rnd{\vec{r},\hat{\vec{\Omega}}} =  \frac{1}{4\pi\Lambda}\phi^{0}\rnd{T\rnd{\vec{r}}} - \frac{\beta\rnd{T\rnd{\vec{r}}}}{4\pi \abs{\vec{v}}} \phi^{0}\rnd{T\rnd{\vec{r}}},
\end{equation}
where the terms on the left are the streaming and collision operators, respectively, with source terms on the right hand side. We have included temperature dependent notation $\rnd{T\rnd{\vec{r}}}$ in Eq.~(\ref{eq:firstOrderRearrange}) as a reference point to show the terms where temperature is coupled in our approach. For much of the remainder of this section, we drop the independent variables $(\vec{r},\hat{\vec{\Omega}})$.

The vacuum and reflecting boundary conditions for this transport equation are defined as:
\begin{equation*}
\psi\left(\vec{r}_{b}\right) = 
\begin{cases}
\psi^{\textrm{vac}}\left(\vec{r}_{b},\hat{\vec{\Omega}}\right), \hat{\vec{\Omega}}\cdot \bar{\vec{n}}_{b} < 0 \\
\psi^{\textrm{ref}}\left(\vec{r}_{b},\hat{\vec{\Omega}}_{r}\right), \hat{\vec{\Omega}}\cdot \bar{\vec{n}}_{b} < 0 
\end{cases},
\end{equation*}
here, $\bar{\vec{n}}_{b}$ is the outward unit normal at a point $\vec{r}_{b} $ on the boundary. In neutron transport space, $\psi^{\textrm{vac}}$ implies a vacuum boundary ($\psi(\vec{r}_{b},\hat{\vec{\Omega}})=0$). However, in this implementation, we leverage the vacuum boundary as an adiabatic boundary to specify an incident source of phonons, e.g., $\psi(\vec{r}_{b},\hat{\vec{\Omega}}) = \phi^{0}\rnd{\vec{r}_{b}}$. The reflective angle $\hat{\vec{\Omega}}_{r} $ in $\psi^{\textrm{ref}}$ is 
\begin{equation*}
\hat{\vec{\Omega}}_{r} = \hat{\vec{\Omega}} - 2\left(\hat{\vec{\Omega}}\cdot \bar{\vec{n}}_{b}\right)\bar{\vec{n}}_{b}.
\end{equation*}
The derivation of the SAAF form of the phonon BTE is based on a straightforward algebraic technique. From a computational perspective the SAAF formulation is advantageous as the full angular flux intensity is the unknown. Upwinding is a common numerical discretization of advection operators in partial differential equations, and is frequently used with the traditional first-order form of the Boltzmann transport equation, particularly in one and two spatial dimensions. The (primarily) lower triangular linear algebra problem in each quadrature direction is solved via a ``transport sweep'' - forward propagation of information through spatial cells in the direction of particle travel. The SAAF equations have an elliptic streaming term, and the resulting linear algebraic equations for the intensity in each angular ordinate are then solved via preconditioned GMRES methods on unstructured spatial grids -- this is the primary numerical advantage to using the SAAF equations; \texttt{boomerAMG}-preconditioned GMRES is massively parallelizable. In contrast, sweeping algorithms (especially on unstructured grids in 3D) are challenging to implement in parallel; the development of an efficient parallel sweep algorithm is an active area of interest in the neutron transport community~\cite{Adams2019}. The use of reflecting boundary conditions is easier with the availability of the full angular flux, as the incoming and outgoing directions are coupled in the same manner as the first order form of the transport equation~\cite{Morel}. Through the application of a continuous finite element (CFEM) spatial discretization, the matrices are symmetric positive definite (SPD), which allows for the use of solution techniques such as the preconditioned Krylov family of solvers. To obtain the SAAF form of the BTE, a simple algebraic approach is followed, solving Eq.~(\ref{eq:firstOrderRearrange}) for the angular intensity:
\begin{equation*}
\psi = \frac{\phi^{0}}{4\pi} - \frac{\Lambda \beta \phi^{0}}{4\pi\abs{\vec{v}}} - \Lambda \hat{\vec{\Omega}}\cdot \nabla \psi,
\end{equation*}
and substitute this expression back into the streaming term in Eq.~(\ref{eq:firstOrderRearrange})
\begin{equation}
\hat{\vec{\Omega}}\cdot \nabla \left( \frac{\phi^{0}}{4\pi} - \frac{\Lambda \beta \phi^{0}}{4\pi\abs{\vec{v}}} - \Lambda \hat{\vec{\Omega}}\cdot \nabla \psi   \right) +\frac{1}{\Lambda}\psi = \frac{\phi^{0}}{4\pi \Lambda}  - \frac{\beta \phi^{0}}{4\pi \abs{\vec{v}}},
\end{equation}
rearranging yields the SAAF form of the phonon BTE
\begin{equation}
\label{eq:SAAFRattlesnake}
-\hat{\vec{\Omega}}\cdot \nabla \sq{\Lambda\hat{\vec{\Omega}}\cdot \nabla \psi} + \frac{1}{\Lambda}\psi =
\frac{1}{4\pi} \sq{\frac{\phi^{0}}{\Lambda} - \hat{\vec{\Omega}}\cdot \nabla \phi^{0} - \frac{\beta \phi^{0}}{\abs{\vec{v}}} + \hat{\vec{\Omega}}\cdot \nabla  \frac{\Lambda\beta\phi^{0}}{\abs{\vec{v}}}}.
\end{equation}
We use the discrete ordinates method~\cite{LewisMiller} to discretize the angular variable. In the discrete ordinates approach, we represent the independent variable $\hat{\vec{\Omega}}$ by a discrete set of directions, $\{\hat{\vec{\Omega}}_{m}, m = 1,\ldots,M\}$. The functions of $\hat{\vec{\Omega}}$ are represented only by their values at each of their directions on the spatial mesh, e.g.,
\begin{equation}
f\rnd{\vec{r}, \hat{\vec{\Omega}}} \rightarrow f\rnd{\vec{r},\hat{\vec{\Omega}}_{m}} \equiv f_{m}\rnd{\vec{r}},\qquad m=1,\ldots,M.
\end{equation}
Consider the angular flux intensity as a function of the direction variable $\hat{\vec{\Omega}}$. There exists a set of complete orthogonal functions in $\hat{\vec{\Omega}}$, much like Legendre polynomials; these are the spherical harmonics functions. To compute the total scalar flux in a group at the end of an iteration, the angular flux intensity is integrated over direction using  Eq.~(\ref{eq:zeroMoment}). However, we make the approximation that the angular flux intensity $\psi(\vec{r},\hat{\vec{\Omega}})$ can be represented by these spherical harmonics functions (termed $Y$) and we expand the angular flux in terms of these functions
\begin{equation}
    \psi\rnd{\vec{r},\hat{\vec{\Omega}}} \approx \frac{1}{4\pi} \sum_{k=0}^{\infty}\sum_{n=-k}^{n=k} \psi_{k,n}\rnd{\vec{r}}Y_{k,n}\rnd{\hat{\vec{\Omega}}},
\end{equation}
where
\begin{equation}
    \psi_{k,n}\rnd{\vec{r}} = \int_{4\pi} Y^{*}_{k,n}\rnd{\hat{\vec{\Omega}}}\psi\rnd{\vec{r},\hat{\vec{\Omega}}}d\Omega.
\end{equation}
This approach involves expanding the angular dependence of the flux intensity in a finite series of spherical harmonics $Y_{l,m}(\hat{\vec{\Omega}}) = Y_{l,m}\rnd{\theta,\varphi}$, a familiar sight in quantum mechanics. Coupling for the angular flux intensity and the transport flux exists in the angular moments alone
\begin{equation*}
     \phi_{l,n}\rnd{\vec{r}}=
     \frac{1}{4\pi} \sum_{l=0}^{L} \sum_{n=-l}^{l}\sum_{m=1}^{M}w_{m}\psi_{m}\rnd{\vec{r}}Y^{*}_{l,n}\rnd{\hat{\vec{\Omega}}_{m}}Y_{l,n}\rnd{\hat{\vec{\Omega}}_{m}},
\end{equation*}
here, $L$ is the truncated spherical harmonics, $l$ is the degree of spherical harmonics and $n$ is the order,  and weights sum to $2\pi$ in 2D
\begin{equation*}
\label{eq:sumWeights}
    \sum_{m=1}^{M} w_{m} = 2\pi.
\end{equation*}
The total heat flux $\vec{q}\rnd{\vec{r}}$ is the sum of the first angular moments ($\phi_{1,0}$ and $\phi_{1,1}$ for two spatial dimensions) over all spectral groups and polarizations
\begin{equation*}
\label{eq:firstMoment}
\vec{q}\rnd{\vec{r}} =\sum_{p} \int_{\eta} \int_{4\pi}\psi\rnd{\vec{r},\hat{\vec{\Omega}}} \hat{\vec{\Omega}}d\Omega d\eta,
\end{equation*}
and has units of $\textrm{W}\cdot\textrm{m}^{-2}$. In this way we solve the transport equation for each angular direction $\hat{\vec{\Omega}}_{m}$. At the end of each iteration, the numerical integration is performed to represent the scalar flux with these angular moments. Then, $\phi^{\textrm{T}}$ is used to determine the local temperatures in the transport system, and are then passed into the equilibrium radiance relations. Now that we have represented the angular flux intensity by this quadrature, the discrete form of Eq.~(\ref{eq:SAAFRattlesnake}) for any direction $\hat{\vec{\Omega}}_{m}$ with an associated weight $w_{m}$ in the angular quadrature set is:
 \begin{multline}
\label{eq:snSAAF}
    -\hat{\vec{\Omega}}_{m}\cdot \nabla \sq{\Lambda\hat{\vec{\Omega}}_{m}\cdot \nabla \psi^{\rnd{\ell+1}}_{m}} +\frac{1}{\Lambda}\psi^{\rnd{\ell+1}}_{m} =
    \\
    \frac{1}{4\pi} \sq{\frac{\phi^{0,\rnd{\ell}}}{\Lambda} - \hat{\vec{\Omega}}_{m}\cdot \nabla \phi^{0,\rnd{\ell}} - \frac{\beta^{\rnd{\ell}} \phi^{0,\rnd{\ell}}}{\abs{\vec{v}}} + \hat{\vec{\Omega}}_{m}\cdot \nabla \frac{\Lambda\beta^{\rnd{\ell}}\phi^{0,\rnd{\ell}}}{\abs{\vec{v}}}},
\end{multline}
where we have assigned iteration indices to Eq.~(\ref{eq:snSAAF}); we solve for the value of angular flux intensity, $\psi^{\rnd{\ell+1}}$ using sources from the previous iteration $\rnd{\ell}$. This resembles source iteration, a well known solution technique in neutron transport~\cite{LewisMiller,accelAdams,LarsenGrey}. This procedure is outlined in Algorithm~\ref{alg::coupledSI}.

The transport calculation assumes that all of the spectral groups are decoupled; each system of equations is solved separately and no information is passed between them. We have demonstrated this capability previously with Rattlesnake~\cite{HarterANS2016,HarterJHT2018}. It is possible to solve each system of equations separately and then compute an average material temperature for the domain. However, this only considers isolated group contributions to temperature and heat flux. Explicit simulation of group-to-group phonon scattering foregoes the SMRT approximation, is computationally demanding, and is an active research area~\cite{mfMinnich,Srivastava}; this work does not consider explicit scattering processes.

The physics dictate that phonon modes are coupled as they become excited through phonon collisions and temperature changes. The amount of heat flowing through each phonon mode in the system must `see' the others -- even forsaking an explicit collision model, individual modes must feel the effects of the material temperature; Bose-Einstein statistics assert this, and the phonon groups must be globally coupled in some manner. A material temperature is determined by prescribing a temperature gradient between spatial boundaries, giving rise to blackbody phonon emission into the material. Phonon scattering in each of the groups are driven by this temperature gradient, but again no coupling exists. This decoupled approach fails to consider the temperature in the Bose-Einstein statistics which must be extended to all phonon modes -- the true equilibrium distribution is an integration coupling all spectral zones. Temperature is defined by the phonon phase space distributions, spanning all modes and polarizations. Phonons are coupled via the internal energy of the crystal system, and we can leverage this detail to derive a set of self-consistent equations where an average material temperature is shared between each of the discrete phonon groups.

The total phonon radiance of the system is defined as
\begin{equation}
\label{eq:totalRadiance}
    \Phi^{00}=\Phi^{00}\rnd{T\rnd{\vec{r}}} = \sum_{p} \int_{\eta} \frac{\hbar \omega\abs{\vec{v}} \mathbb{D}}{\exp\left[\frac{\hbar \omega}{k_\text{B}T\rnd{\vec{r}}}\right]-1}d\eta,
\end{equation}
which is just the contribution of $\phi^{0}$ over all modes. 
We make the assumption that the total equilibrium radiance $\Phi^{00}$ can be related to the total transport scalar flux in the domain, such that the two must be in radiative equilibrium. This treatment uses the Bose-Einstein distribution as the phonon source for each of the spectral groups as opposed to the transport scalar flux; we still perform an integration of the angular intensity at the end of each linear iteration to obtain transport scalar flux. Equating $\Phi^{00}$ to the sum of the transport flux, we have
\begin{equation}
    \label{eq:tempDer0}
    \Phi^{00}\rnd{\vec{r}} = \sum_{p}\sum_{\eta} \phi^{\mathrm{T}} = \sum_{p}\sum_{\eta} \frac{\hbar \omega_{p,\eta}\left| \vec{v}_{p,\eta}\right| \mathbb{D}_{p,\eta} }{\exp\sq{\frac{\hbar \omega_{p,\eta}}{k_\text{B}T\rnd{\vec{r}}}} - 1},
\end{equation}
where the $\rnd{p,\eta}$ index denotes a double summation over the polarizations and spectral property indices. Now we may carry out the solution for $T\rnd{\vec{r}}$, which involves summations over the angularly dependent material properties (index $m$) and group index $g$, both described in proceeding Sec.~\ref{sec:transportProps}.
\begin{equation}
    \label{eq:tempDer1}
    T\rnd{\vec{r}} = \sum_{m,g} \frac{\hbar \overline{\omega}_{m,g}}{k_{\textrm{B}} \ln \sq{\frac{\hbar\overline{\omega_{m,g} v_{m,g}\mathbb{D}}_{m,g}}{\phi^{\mathrm{T}}_{g}\rnd{\vec{r}}}+1}}.
\end{equation}
Equation~(\ref{eq:tempDer1}) is used to compute an average temperature for the entire spatial domain. $T\rnd{\vec{r}}$ is what couples the phonon groups together -- they now depend on a global average temperature. The resulting value of $T\rnd{\vec{r}}$ is used to compute a new value of $\phi^{0}$ using  Eq.~(\ref{eq:radianceDefinition}).

Equation~(\ref{eq:snSAAF}) in the discrete ordinate method generates a linear system of equations arising from the spatial discretization of an elliptic operator, and is solved for angular intensity $\psi_{p,\eta,m}$ in each quadrature direction. This means that software for solving the diffusion approximation to transport can be exploited to implement acceleration techniques, which will be part of a future study. The Galerkin weak form of Eq.~(\ref{eq:snSAAF}) using weighted residual formulation with CFEM is then solved with Rattlesnake. The derivation of the weak form of the SAAF-BTE for phonons is similar to that of the weak form of the SAAF-BTE for neutrons and can be followed in the Rattlesnake theory~\cite{rsm}. Algorithm~\ref{alg::coupledSI} is outlines the technique used to solve the system of equations.
\\
\begin{algorithm}[H]
\DontPrintSemicolon
\LinesNumbered
initialize $\phi^{0}\rnd{\vec{r}_{b}}$,\,$\psi^{\rnd{\ell}}_{m}\rnd{\vec{r}}$,\,$\phi^{\textrm{T},\rnd{\ell}}\rnd{\vec{r}}$,\,$\phi^{0,\rnd{\ell}}\rnd{\vec{r}}$,\,$\Phi^{00, \rnd{\ell}}\rnd{\vec{r}}$, $\Phi^{\textrm{T},\rnd{\ell}}$ $T^{\rnd{\ell}}\rnd{\vec{r}}$\;
set convergence tolerance $\varepsilon$\;
    {\For{$g=1$ to $G$}{
        \For{$m=1$ to $M$}{
        solve SAAF-BTE (Eq.~\ref{eq:snSAAF}) for $\psi_{m}^{\rnd{\ell+1}}$\;
        }
    compute $\phi^{\textrm{T},\rnd{\ell+1}}\rnd{\vec{r}}$ and solve for $T^{\rnd{\ell+1}}\rnd{\vec{r}}$\;
    compute $\phi^{0}\rnd{T^{\rnd{\ell+1}}}$\;
    }
}
compute $\Phi^{00,\rnd{\ell+1}}$, $\Phi^{\textrm{T},\rnd{\ell+1}}$, $\beta^{\rnd{\ell+1}}$\;
\uIf{$\left| \frac{\phi^{\textrm{T},\rnd{\ell+1}}\rnd{\vec{r}} - \phi^{\textrm{T},\rnd{\ell}}\rnd{\vec{r}}}{\phi^{\textrm{T},\rnd{\ell+1}}\rnd{\vec{r}}}\right| < \varepsilon$}{break}
\uElse{go to 3}
\caption{Temperature coupled iteration}
\label{alg::coupledSI}
\end{algorithm}

\subsection{Phonon Transport Properties of Silicon}\label{sec:transportProps}
For the purposes of demonstrating and testing the use of Rattlesnake to simulate spectrally-resolved phonon transport we use silicon as a model material system. The thermal conductivity of Si is isotropic and well characterized, and there are a wealth of experimental and computational data against which to compare our model. Silicon has cubic crystal symmetry; to simplify our model and to enable us to use a standard set of transport ordinates akin to neutron transport (which does not have cubic symmetry), we simplify the transport spectrum to make it spherically symmetric. The discussion that follows describes first the process we used to compute \emph{ab initio} the phonon properties of Si, and the metric we have developed for grouping the phonon modes of Si into a discrete set of transport groups with aggregated transport properties.

\subsubsection{Calculation of phonon properties}
The frequency and group velocity of phonons with a given wave vector and polarization are computed from the dispersion relation obtained from the dynamical matrix of second order force constants. The probability of three phonon scattering processes, the dominant intrinsic scattering process at high temperature, are computed from the matrix of third order stiffness interactions. 
The \nth{2} and \nth{3} order force constants for Si at 0 K were obtained from the \emph{AlmaBTE} database~\cite{Carrete_2017}. In this data set the \nth{2} order force constants were computed via the frozen phonon method as implemented using \emph{phonopy}~\cite{phonopy} using a large $5\times5\times5$ supercell in order to accurately capture the contributions from long ranged interatomic interactions. The \nth{3} order force constants were calculated using a $5\times5\times5$ supercell using the package \emph{thirdorder.py}~\cite{ShengBTE_2014} and considered all three-atom interactions of atoms less than five neighbours removed from one another.
The complexity of the unit cell does not directly affect computational cost, but it can have indirect influence on the computational cost. The main source of computational expense comes from how finely one resolves the phonon spectrum, which depends on the span of frequencies and mean free paths. However, the CPU time is impacted by the acoustic thickness of a particular angle-group; if the unit cell has many optical modes with low velocity and short lifetime, these diffusive modes will take longer to converge. We discuss this further in a proceeding section. The full details of the DFT calculations are given in reference~\cite{ShengBTE_2014}.

Starting with these stiffness matrices we used \emph{AlmaBTE} to compute the full phonon dispersion and group velocities on a $24\times24\times24$ Brillouin zone mesh. The 0 K velocities and frequencies were assumed to be unchanged at higher temperatures. The phonon-phonon scattering times were computed on the same mesh at a temperature of 300 K. All of the phonon modes in the Brillouin zone are represented in the transport groups, albeit in an approximate way. In this work, this was done to make the transport spectrum spherically symmetric; however it is no extra work to include all the modes in the Brillouin zone in a non-approximate way, and to model anisotropic materials. For this, one would need to use a set of ordinates which match the crystal symmetry, and we would not be able to decouple the transport group's properties and directions.

\subsubsection{Decomposition of phonon spectrum into transport groups}
Discretization of the BTE in Eq.~(\ref{eq:snSAAF}) means solving for the heat flux in a set of transport groups where each group represents the combined contribution to transport from a collection of phonon modes with similar direction of propagation and spectral character. The task is to determine the effective transport properties of each group such that all phonon modes are accounted for, and that the temperature in Eq.~(\ref{eq:tempDer1}) may be inverted.

In discretized form, the angular and spectral distribution of a quantity $A(\hat{\vec{\Omega}},\eta,p)$ of the population is approximated as a sum of piece-wise basis functions:
\begin{equation*}
    \label{eq:Aapprox}
{A\left(\hat{\vec{\Omega}},\eta,p\right) \approx \sum_{m=1}^{N_\Omega} \sum_{g=1}^{N_\eta} A_{m,g,p} \chi_{m}\left(\hat{\vec{\Omega}}\right)\chi_g\left(\eta\right).}
\end{equation*}
Here $m$ is the index over discrete ordinates in $\hat{\vec{\Omega}}$, and $g$ is the index over control points (\emph{groups}) in $p$ and $\eta$. The functions $\chi_m(\hat{\vec{\Omega}})$ and $\chi_g\left(\eta\right)$ are element shape functions --- non overlapping boxcar functions that satisfy the relationships
\begin{equation*}
\int_{\eta} d\eta \int_{4\pi} d\Omega\, \chi_{m}\left(\hat{\vec{\Omega}}\right)\chi_{g}\left(\eta\right) = w_m \Delta\eta_g,
\end{equation*}
where $\Delta\eta_g$ is the width of the $g$-th spectral group, and
\begin{equation*}
\int_{\eta} d\eta \int_{4\pi} d\Omega~ \chi_{m}\left(\hat{\vec{\Omega}}\right)\chi_g\left(\eta\right)~\chi_{m'}\left(\hat{\vec{\Omega}}\right)\chi_{g'}\left(\eta\right) = \begin{cases}
	w_m^2 \Delta\eta_g^2, & \text{if $m=m'$ and $g=g'$}\\
    0, & \text{otherwise}
\end{cases}.
\end{equation*}
The transport group quantities $A_{m,g}$ represent the combined contribution to transport from all of the phonon modes that fall into the $m,g$-th group. To perform this calculation we define the group integration operator
\begin{equation*}
    G_{m,g}\left\{A\left(\hat{\vec{\Omega}},\eta,p\right)\right\} = \sum_d \int_{\eta} d\eta \int_{4\pi} d\Omega\,A\left(\hat{\vec{\Omega}},\eta,p\right)~ \chi_{m}\left(\hat{\vec{\Omega}}\right)\chi_{g}\left(\eta\right).
\end{equation*}
Using this operator we define the groups' mean free path as the average of the mean free paths of the phonon modes in the group weighted by their contribution to the phonon flux
\begin{equation*}
    \Lambda_{m,g}=\frac{G_{m,g}\left\{\Lambda \phi^{0}\rnd{\overline{T}}\right\}}
    {G_{m,g}\left\{\phi^{0}\rnd{\overline{T}}\right\}}.
\end{equation*}
Where $\overline{T}$ is the expected average temperature of the system to be simulated. Similarly, for computing radiance from temperature we define:
\begin{eqnarray}
    \phi^{0}_{m,g}\rnd{\overline{T}} &=& \frac{1}{w_m} G_{m,g}\left\{\phi^{0}\rnd{\overline{T}}\right\} =
    \frac{1}{w_m} G_{m,g}\left\{\frac{~\hbar\omega\left|\vec{v}\right|\mathbb{D}}{\exp\rnd{\frac{\hbar\omega}{k_B T}}-1}\right\},\nonumber\\
\phi^{0,'}_{m,g}\rnd{\overline{T}} &=& \frac{1}{w_m} G_{m,g}\left\{\frac{d\phi^{0}\rnd{\overline{T}}}{dT}\right\},\nonumber\\
\phi^{0,''}_{m,g}\rnd{\overline{T}} &=& \frac{1}{w_m} G_{m,g}\left\{\frac{d^{2}\phi^{0}\rnd{\overline{T}}}{dT^2}\right\}.\nonumber
\end{eqnarray}
The only material properties in the discretized BTE (Eq.~(\ref{eq:snSAAF})) are the groups' effective mean free path and radiance, however the transport simulation needs to be able to compute the local temperature and so it also requires the weighting terms to obtain this from the sum over fluxes. To compute the energy in a group from the angular flux, we define the group velocity as the averaged velocity weight by mode energy  
\begin{equation*}
    \overline{v}_{m,g}=\frac{G_{m,g}\left\{\phi^0\rnd{\overline{T}}\right\}}
    {G_{m,g}\left\{\frac{\phi^0\rnd{\overline{T}}}{\left|\vec{v}\right|}\right\}}.
\end{equation*}
For the local temperature calculation we need to know the total of the product of frequency, density of states and velocity
\begin{equation*}
    \overline{\omega v \mathbb{D}}_{m,g} = G_{m,g} \left\{\omega\left|\vec{v}\right|\mathbb{D} \right\},
\end{equation*}
and the average frequency
\begin{equation*}
    \overline{\omega}_{m,g}=\frac{G_{m,g}\left\{\omega\mathbb{D}\right\}}{G_{m,g}\left\{\mathbb{D}\right\}}.
\end{equation*}

\begin{figure}[ht]
\centering
    \includegraphics[scale=.60]{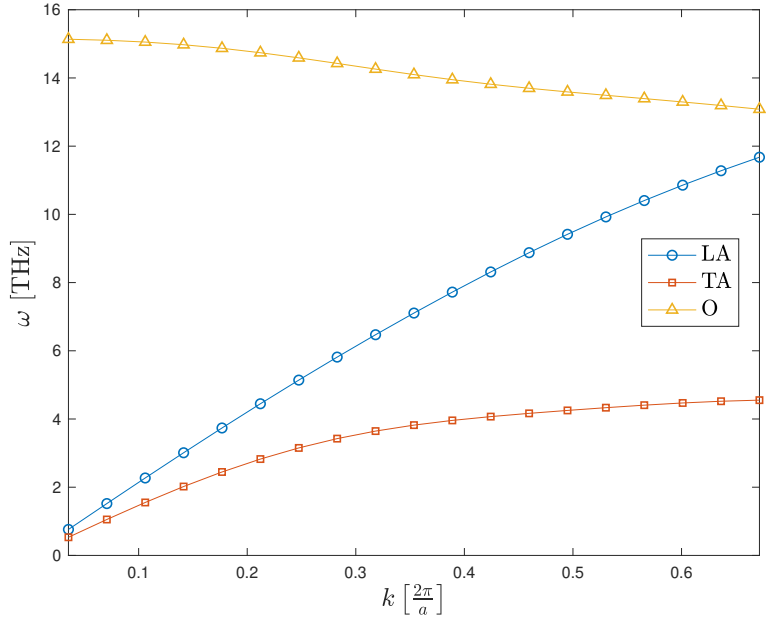}
    \caption{\label{fig:dispersion} Dispersion relation in silicon.}
\end{figure}

The approach developed up to this point is quite general; we have not yet specified a choice of spectral parameterization, and the method can be applied to anisotropic materials if one uses a set of transport ordinates with the appropriate symmetry. In the remainder of the manuscript we limit ourselves to one choice for parameterization of the dispersion in Si, in which we approximate the dispersion relation to be spherically symmetric enabling us to use a standard set of non-symmetric transport ordinates. In this approach we used just the dispersion for Si along the (100) crystal direction, and we collapsed the three optical branches and the two transverse acoustic branches into a single branch with averaged frequency as shown in Fig.~\ref{fig:dispersion}. The dispersion was parameterized in wave \emph{number}, $k$, and this condensed (100) dispersion was used as the dispersion along all directions. The density of states was defined as 
\begin{equation*}
    \mathbb{D}\rnd{\hat{\vec{\Omega}},k,p}_{m,g}  = \frac{3 d_p k^2}{\pi k_{\mathrm{max}}^3 a_{\mathrm{Si}}^3},
\end{equation*}
where $d_p$ is the degeneracy that accounts for the merged branches and $a_{\mathrm{Si}}$ is the lattice parameter of the Si unit cell. The integral of this accounts for all the phonon modes of Si
\begin{equation*}
    \sum_{p=1}^3\int_{4\pi}d\Omega \int_0^{k_{\mathrm{max}} }dk \frac{3 d_p k^2}{\pi k_{\mathrm{max}}^3 a_{\mathrm{Si}}^3} = \frac{3\times 8}{a_{\mathrm{Si}}^3} = 3 N_v,
\end{equation*}
The approximated dispersion was split into between three and 20 groups in $k$ of equal size.

\begin{table*}
\caption{\label{tab:materialProps}Silicon material properties.}
\begin{ruledtabular}
\begin{tabular}{ccccccc}
G & $\Lambda$ [nm] & $\overline{\omega}\cdot 10^{13}\, \sq{s^{-1}}$ & $\overline{v}\,\sq{\textrm{m}\cdot \textrm{s}^{-1}}$ & $\overline{\mathbb{D}}\cdot 10^{27}\, \sq{\textrm{m}^{-3}}$ & $\overline{\tau}\,\sq{\textrm{ps}}$ & $\overline{\omega v \mathbb{D}}\,\sq{\textrm{m}^{-2}\cdot \textrm{s}^{-2}}$ \\
\hline
1 (LA) & 3120 & 1.49 & 8079 & 0.0736 & 386  & $12.9\cdot 10^{42}$ \\
2 (LA) & 155  & 4.25 & 6926 & 0.515  & 22.4 & $156\cdot 10^{42}$  \\
3 (LA) & 28.7 & 6.46 & 5152 & 1.4    & 5.6  & $458\cdot 10^{42}$  \\
4 (TA) & 898  & 0.99 & 5076 & 0.147  & 177  & $9.4\cdot 10^{42}$  \\
5 (TA) & 79.5 & 2.3  & 2063 & 1.03   & 38.5 & $42.2\cdot 10^{42}$ \\
6 (TA) & 11.1 & 2.75 & 721  & 2.8    & 15.4 & $51.8\cdot 10^{42}$ \\
7 (O) & 3.15 & 9.43 & 786  & 0.22    & 4    & $24.9\cdot 10^{42}$ \\
8 (O) & 3.8  & 8.92 & 1651 & 1.55    & 2.3  & $219\cdot 10^{42}$  \\
9 (O) & 4.58 & 8.43 & 1126 & 4.2     & 4.1  & $397\cdot 10^{42}$  \\
\end{tabular}
\end{ruledtabular}
\end{table*}

\begin{figure}[ht]
    \centering
    \includegraphics[scale=.60]{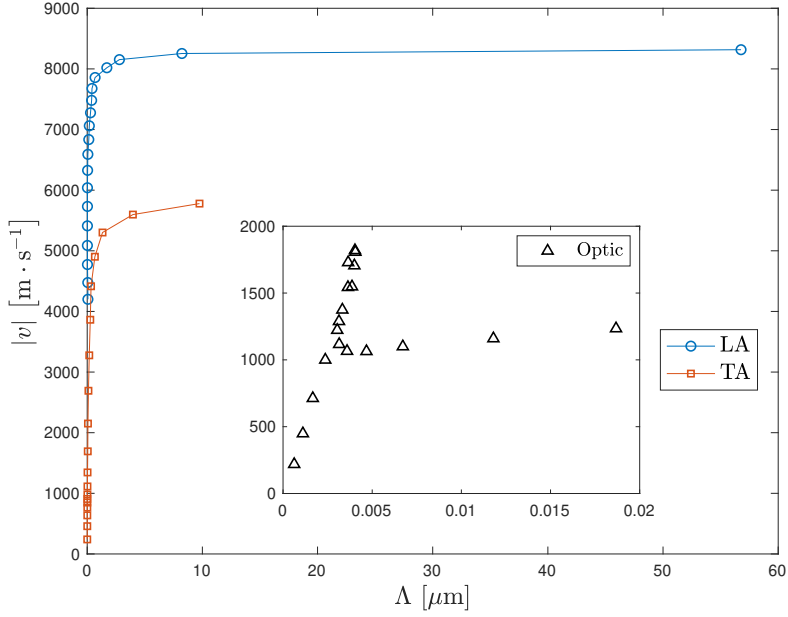}
    \caption{\label{fig:mfp-velocity} Mean free paths for dispersion branches vs. group velocity. Units are consistent for inset figure.}
\end{figure}

Using these data, we can compute radiant emission sources on the boundaries, driven by a prescribed temperature boundary condition,
\begin{equation}\label{eq:boundarySource}
\phi^{0}_{p,\eta}\left(T\left(\vec{r}_{b}\right)\right) = \frac{\hbar \omega_{p,\eta} \left|\vec{v}_{p,\eta}\right| \mathbb{D}_{p,\eta}} {\exp\sq{ \frac{\hbar\omega_{p,\eta}}{k_\text{B} T\left(\vec{r}_{b}\right) }} - 1}.
\end{equation}
Equation~(\ref{eq:boundarySource}) is used by the vacuum condition at a boundary $\vec{r}_{b}$ to supply an incident source of phonons for a given group and polarization. We compute an effective thermal conductivity along a direction by taking the ratio of the total heat flux to the end-to-end temperature gradient (which includes boundary effects) in the system
\begin{equation}
\label{eq:kappa}
\kappa_{x}\left(\vec{r}\right) = \frac{1}{\left[T\left(x_\textrm{L}\right) - T\left(x_\textrm{R}\right)\right]} \frac{1}{L_y L_z} \int \vec{e}_{x} \cdot \vec{q}\left(\vec{r}\right) d^{3}\vec{r}.
\end{equation}

\section{Results}
\begin{figure}
	\centering
 		\includegraphics[scale=.35]{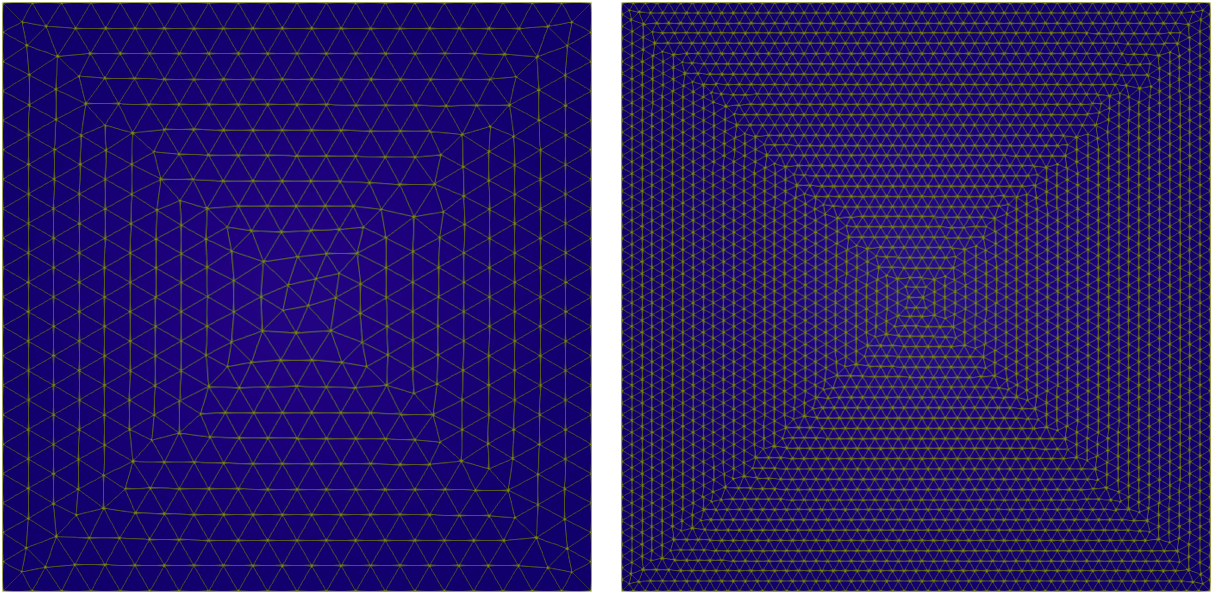}
\caption{\label{fig:meshes}Coarse (left, 926 elements) and fine (right, 5770 elements) spatial meshes.}
\end{figure}

We performed spectrally resolved phonon transport simulations in two-dimensional planes of silicon of varying geometric size. We report heat flux, thermal conductivity, and the equilibrium temperature distribution. We used full phonon dispersion and density of states computed at room temperature from \emph{ab initio} DFT simulations. Spatial domain sizes varied from 10 nm to 10 $\mu$m and were spatially discretized using coarse (C) and fine (F) triangular finite element meshes (Fig.~\ref{fig:meshes}). We employed $S_{4}$, $S_{8}$, and $S_{16}$ Gauss-Legendre angular quadratures. We simulate a 1 K temperature gradient along the $x$-axis, with boundary temperatures of $T_\textrm{L} = 301\, \text{K}$, $T_\textrm{R} = 300\, \text{K}$. Reflecting conditions are placed on remaining boundaries. We use AMG-preconditioned GMRES~\cite{boomeramg,Saad1,Saad2} to solve the linear system of equations, with convergence criteria set to $\epsilon = 10^{-8}$. The selected four cases we discuss in this section were all simulated using $S_{8}$ quadrature with the fine spatial mesh.

\begin{figure}
	\centering
 		\includegraphics[scale=.60]{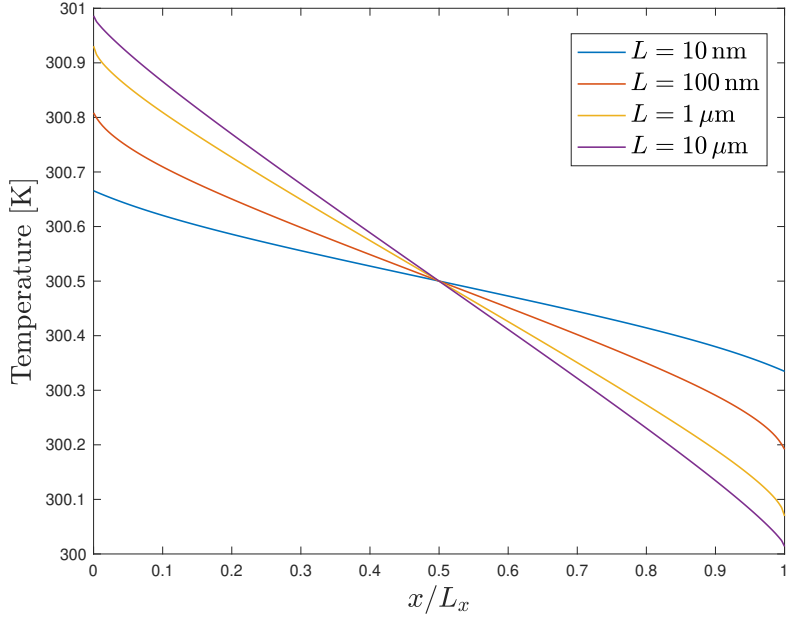}
\caption{\label{fig:temperatureLines}Centerline temperature distribution for four cases. Temperature slips at the boundaries are observed, and their magnitude diminishes in proportion to increasing domain size, as phonon boundary emission sources become further separated.}
\end{figure}

In centerline temperature profiles (Fig.~\ref{fig:temperatureLines}), it is clear that non-equilibrium behavior arises when the domain is small, and this occurs because radiative equilibrium cannot be established. The incident phonon radiance from opposing sides encounter interference due to their proximity -- the distance between hot and cold sources is less than the mean free path of the majority of phonons. This feature is prominent in smaller geometric domain sizes, exacerbated by the presence of ballistic phonons, which may undergo very few collisions before reaching the opposite side of the domain. An equilibrium solution to the temperature distribution exists upon numerical convergence of the simulation as the temperature distribution in each phonon mode is identical. However, if an equivalent transient simulation was conducted, the modal temperature distributions may not be identical. The temperature profiles produced from this work may provide a basis for the benchmarking of temperature distributions in molecular dynamics (MD) simulations, which rely on the careful selection of fitting parameters; the temperature profiles computed by our method may aid in the accurate fitting of these parameters~\cite{Chen2010}.

\begin{figure}
\centering
\includegraphics[scale=.60]{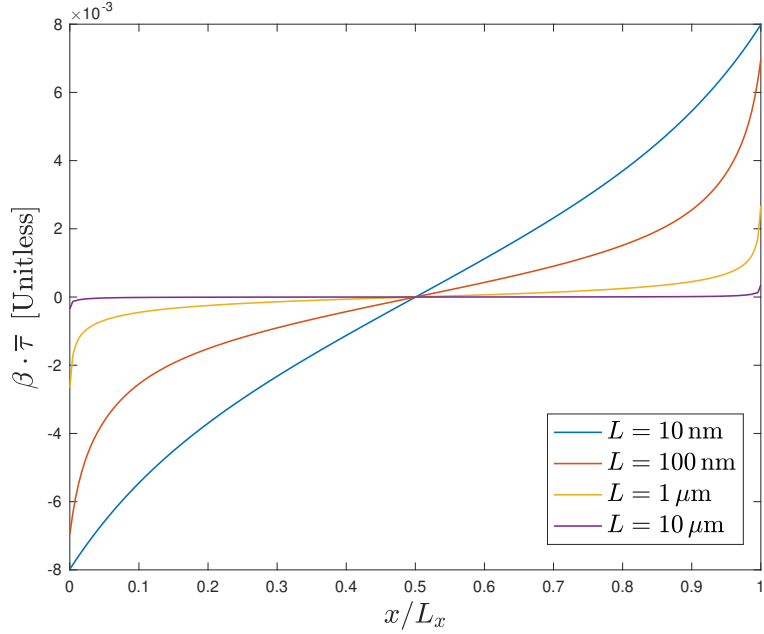}
\caption{\label{fig:betaLines}$\beta\cdot\overline{\tau}$ distributions along the $x$-axis for four domain sizes. As domain size increases, the behavior of $\beta$ in the bulk tends toward zero as the total radiance ($\Phi^{00}$) and total flux ($\Phi^\textrm{T}$) come into balance and the influence of the heated boundaries subsides. The left boundary emits phonons at a higher temperature than the right boundary and so locally, $\phi^\textrm{T} > \phi^{0}$. The opposite is true on the right side of the figure, in proximity of the colder phonon source. The influence of $\beta$ is directly affected by the size of the domain; with smaller domain sizes, the emissive sources are in greater competition with each other and radiative equilibrium can not be established.}
\end{figure}

The closure term $\beta$ (Fig.~\ref{fig:betaLines}) provides a glimpse into the non-equilibrium behavior of the simulation at various domain sizes; $\beta$ tightly couples the total phonon radiance ($\Phi^{00}$) and total flux ($\Phi^\textrm{T}$). We have shown $\beta$ multiplied by the average relaxation time $\overline{\tau}$ in order to understand the fraction of total energy exchanged into the bath in one scattering event. The total amount of energy exchanged in $\beta$ between $\Phi^{\textrm{T}}$ and $\Phi^{00}$ is infinitesimal relative to the total energy of the system, but without the presence of $\beta$ conservation is broken, and a parabolic, rather than constant heat flux is observed.

$\Phi^\textrm{T}$ and $\Phi^{00}$ are not equivalent at small domain sizes; this occurs due to the disparity between the boundary emission source and the localized transport flux, and it is no surprise that the majority of the curvature exhibited by $\beta$ exists near the boundaries of Fig.~\ref{fig:betaLines}. This artifact is a direct consequence of the system size, where emitting boundaries are in competition with each other. $\beta$ exhibits a sign change, beginning negative at $T_\textrm{L}$ and turning positive at $T_\textrm{R}$. In cases where $\beta < 0$, $\phi^\textrm{T} > \phi^{0}$; $\phi^\textrm{T}$ is strongly influenced by the boundary conditions, where the incident angular intensity $\psi$ is specified by the phonon radiance relation in Eq.~(\ref{eq:boundarySource}). As phonons flow from hot to cold sides, the sign of $\beta$ tends positive as $\phi^{0} >\phi^\textrm{T}$, after the spatial midpoint. This is because the local radiance $\phi^{0}$ is now stronger than local transport flux $\phi^\textrm{T}$ -- the cold boundary can not overcome the flow of phonons from the hot side, which results in a positive $\beta$, but also elevates the temperature away from the cold boundary shown on the profiles in Fig.~\ref{fig:temperatureLines}. Blackbody phonon emission is very strong in proportion to the transport flux, and although overall convergence is achieved the localized non-equilibrium effect is very apparent, especially in smaller domain sizes. 

\begin{figure}
	\centering
	\includegraphics[scale=.60]{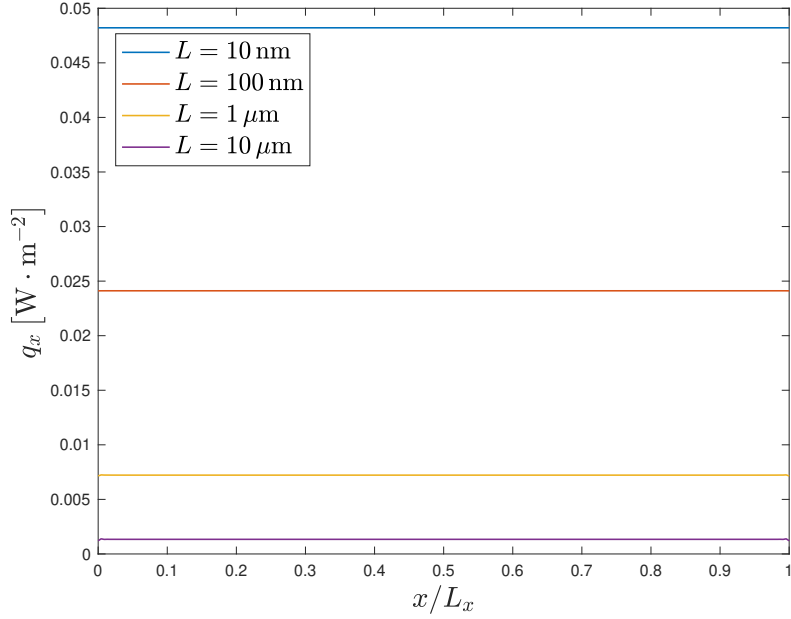}
\caption{\label{fig:heatflux-cl}Centerline total heat flux for the $S_{8,F}$ case. Heat flux is approximately constant across the domain, with very minor fluctuations occurring at the hot and cold emitting boundaries. With increasing domain size, heat flux decreases.}
\end{figure}

\begin{figure}
	\centering
	\includegraphics[scale=.60]{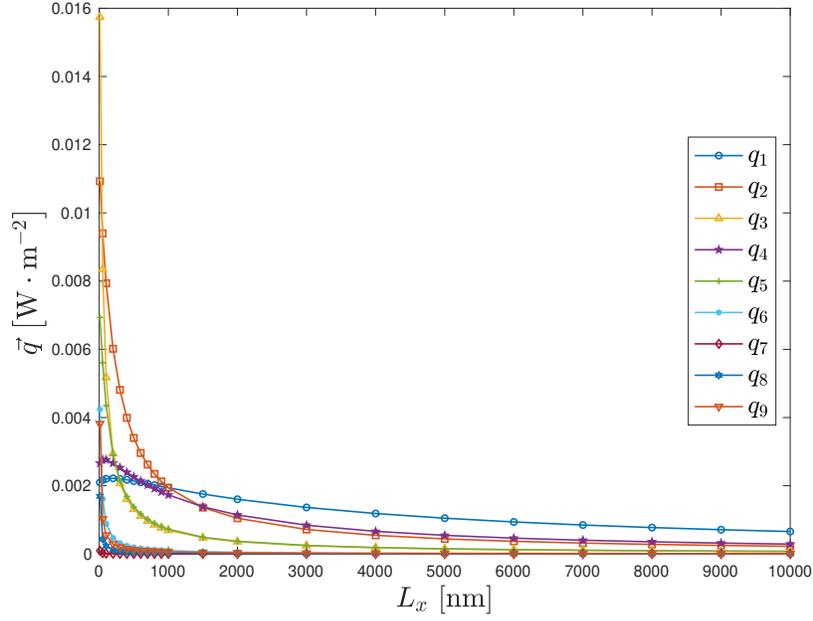}
\caption{\label{fig:heatflux-group}Group heat flux for each geometric domain size for the $S_{8,F}$ case. Groups 1-3 are LA, 4-6 are TA, and 7-9 are O. Diffuse groups (7, 8, 9) always carry low amounts of heat and remain relatively flat independent of domain size.}
\end{figure}

Centerline total heat flux profiles for the $S_{8,F}$ case are shown in Fig.~\ref{fig:heatflux-cl}. As system size increases, heat flux decreases. Because of the flat heat flux profiles observed in all cases, it is clear that including $\beta$ to close the transport equation is a necessity. Without $\beta$, energy leaks out of the system and yields a non-conserved heat flux. However, inspecting group heat flux paints a clear image of which groups carry the most heat in the system. It is, of course, the groups with the largest $\Lambda$, and those which remain the most ballistic over the entire geometric domain range. Figure~\ref{fig:heatflux-group} shows group heat flux for the $S_{8,F}$ case for all domain sizes. It is clear that larger values of $\Lambda_{p,\eta}$ are responsible for a higher heat flux; ballistic phonons carry energy further between collisions. The LA phonons are the dominant carriers due to their higher velocities and mean free path whereas TA and O phonons contribute lesser quantities. Acoustic phonons are generated by atomic displacements moving in phase, which is responsible for their higher velocity. Conversely, optic phonon motion is out of phase and are shorter range carriers. This phenomenon can be traced to Fig.~\ref{fig:mfp-velocity}, where larger magnitudes of the derivatives of the dispersion curves equate to higher wave propagation speed. Heat flux flattens out as geometric size increases, shown in Fig.~\ref{fig:heatflux-group}; in this way we can analyze the degree to which the heat-carrying groups are affected by the change in domain size. If phonon groups were decoupled, the relation between heat flux and domain length would remain fixed in each group, as individual groups are not influenced by the average energy of the domain. However, in a coupled simulation the equilibrium distribution of each phonon channel is impacted by the average material temperature. Results for heat flux and thermal conductivity vary by approximately 3\% between all angular and spatial resolutions. Thus, for a homogeneous material, a moderate angular and spatial resolution is sufficient for accurate results. This will likely change in heterogeneous environments such as porous materials, bulk material with dopants or inclusions (e.g. UO$_{2}$ with Xe bubbles), where ray effects have been observed in grey simulations~\cite{HarterJHT2018}.

\begin{figure}
	\centering
	\includegraphics[scale=.60]{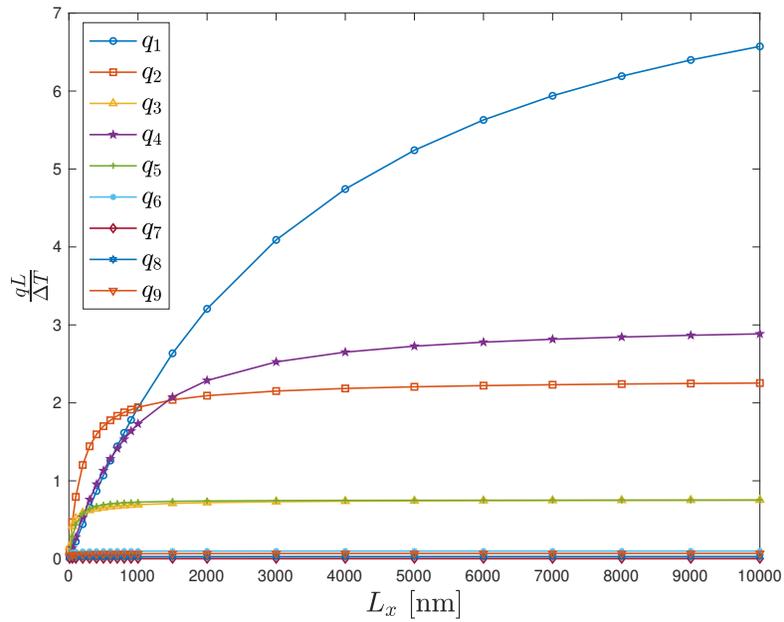}
\caption{\label{fig:heatfluxNormalized}Normalized group heat flux ($S_{8,F}$ case); the ballistic limit is achieved rapidly by most of the phonon groups, with the exception of group 1 (largest $\Lambda$ in the system).}
\end{figure}

Normalized spectral heat flux is shown in Fig.~\ref{fig:heatfluxNormalized}, giving insight into how rapidly each of the groups approach the ballistic limit. It is clear that $q_{1}$ (with $\Lambda_{1} = 3120$ nm) carries the dominant portion of energy throughout the entire range of domain sizes, as it barely begins to approach the asymptotic limit at a length of 10 microns. However, it is overtaken early on by groups 2-5, until the domain sizes reaches about 1000 nm. As expected, the diffuse groups do not contribute appreciably to the overall heat flux.

\begin{figure}
	\centering
	\includegraphics[scale=.60]{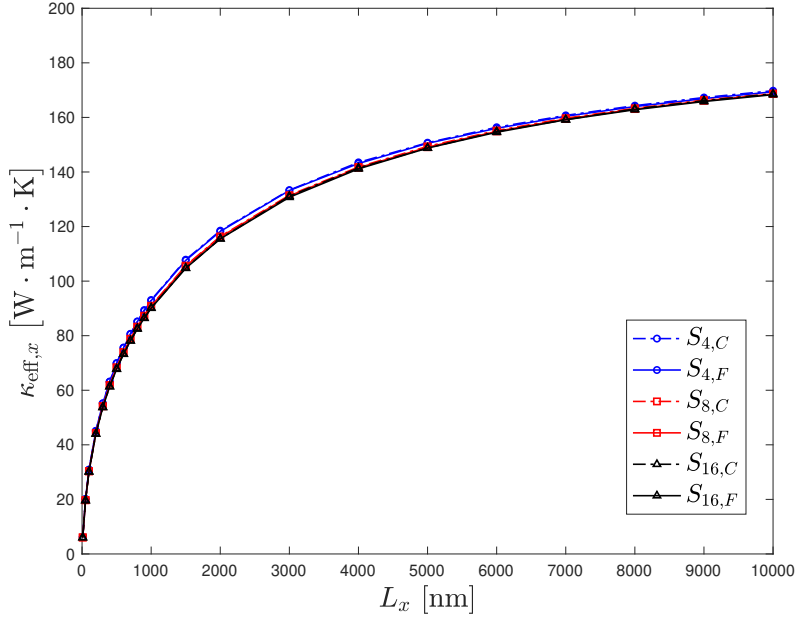}
\caption{\label{fig:kappaPlot}Thermal conductivity for various angular and spatial (F = fine, C = coarse) discretizations up to 10 $\mu$m. The overall difference between the extreme cases is about 3\%.}
\end{figure}

Thermal conductivity for all cases is shown in Fig~\ref{fig:kappaPlot}. The increasing angular refinements reduce heat flux and thermal conductivity by minute amounts, caused by additional information present due to a higher number of streaming directions. The sizes of the domains span 10 nm to 10 $\mu$m and $\kappa_\textrm{eff}$ is proportional to this size. As domain size increases, $\kappa_\textrm{eff}$ asymptotically approaches a bulk value of about $165\,\textrm{W}\cdot\textrm{m}^{-1}\cdot\textrm{K}^{-1}$, and many researchers have reported values for $\kappa_\textrm{eff}$ with variation on the order of 100\%~\cite{Chakraborty2019}. As spatial and angular discretizations change, so do the values of spectral $\kappa_{\textrm{eff}}$. On smaller spatial meshes, $\Lambda_{p,\eta}$ tends to be ballistic per cell, even if they would be diffuse on the global domain. The coarse spatial discretization is acoustically thicker (more diffuse) where the fine spatial discretization is acoustically thinner (more ballistic). It is only as global domain increases in size that these groups become acutely diffuse, and their contribution to total heat flux becomes quite negligible; this trend is shown in Figs.~\ref{fig:heatflux-group} and \ref{fig:heatfluxNormalized}. Table~\ref{tab:O_comparison} contains the contributions from optic phonons to $\kappa_\textrm{eff}$, and they have a stronger effect in smaller global domains. For simulations with large (greater than 1 $\mu$m) domain sizes, omitting optic phonons would cause a dramatic decrease in the amount of iterations required to converge the solution, with negligible loss of accuracy (Table~\ref{tab:O_comparison}). Overall, it is more important to finely resolve spatial cell (finite element) size in comparison to angular resolution.

\begin{figure}
	\centering
	\includegraphics[scale=.60]{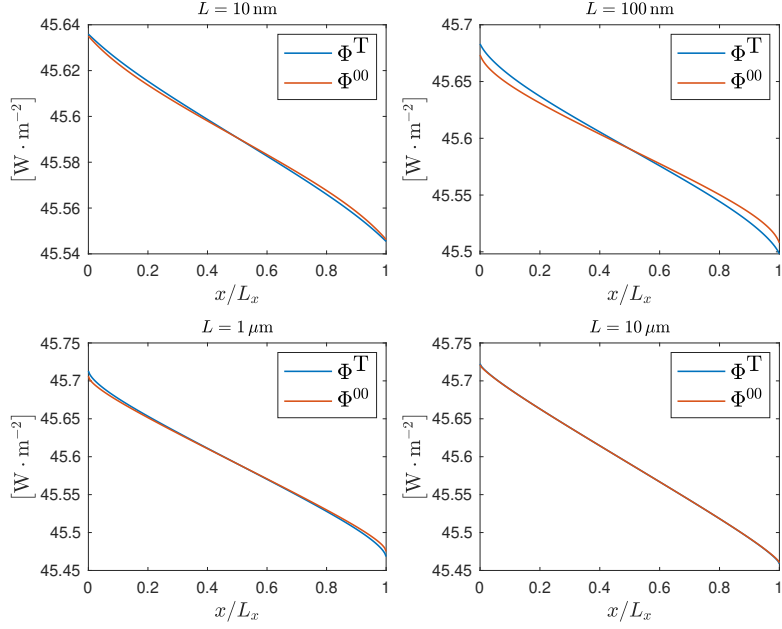}
\caption{\label{fig:fluxAndRadiance}Total radiance ($\Phi^{00}$) and transport flux ($\Phi^\textrm{T}$) comparisons domain sizes between 10 nm and 10 $\mu$m using an $S_{8}$ quadrature. As the domain size increases, the systems relax to a more equilibrate state.}
\end{figure}

Comparisons of $\Phi^{00}$ and $\Phi^\textrm{T}$ are shown in Fig.~\ref{fig:fluxAndRadiance}. The distance between $\Phi^{00}$ and $\Phi^\textrm{T}$ are related to the magnitude of $\beta$, and intuitively, as domain size increases and the magnitude of $\beta$ decreases, $\Phi^{00}$ and $\Phi^\textrm{T}$ should converge. However, for the 100 nm domain this is not the case; the two fluxes are further apart. Looking back to Fig.~\ref{fig:heatflux-group}, there is a slight bump for groups 1 and 4, where heat flux experiences a slight rises rather than a monotonic decrease. Modal dominance of heat flux has an affect on the magnitude between $\Phi^{00}$ and $\Phi^\textrm{T}$ at 100 nm, where the system is still sufficiently far from equilibrium. The same effect is present in domains up to 600 nm, after which the heat flux contribution from group 4 at 700 nm finally becomes less than the heat flux at 50 nm. This lag is caused by phonons in group 1 finally overtaking those from group 4; this transition happens in between 600 and 700 nm, and is observed in Fig.~\ref{fig:heatfluxNormalized}. As the geometric domain increases to the micrometer scale, $\Phi^{00}$ and $\Phi^\textrm{T}$ become increasingly convergent and their respective difference reduces to $<1\%$. These criteria lead to the system being thrust into a state of equilibrium, compared to differences in the distributions on the nanometer scale where the system is far from equilibrium.

\begin{figure}
	\centering
	\includegraphics[scale=.60]{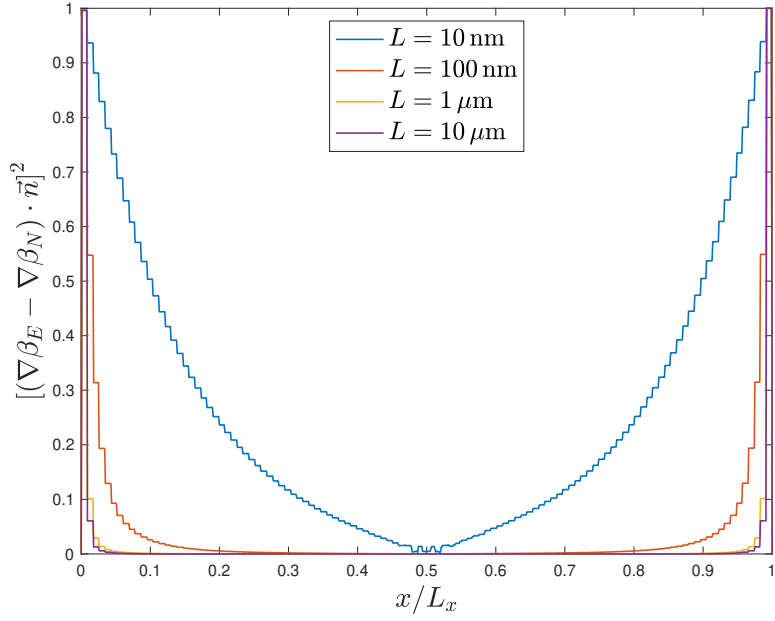}
\caption{\label{fig:AMR}Normalized error from the AMR process. Differences in the solution gradient of $\beta$ for various domain sizes. The strongest gradient jumps are observed near the edges, a result of the fixed boundary condition. For the 10 nm case, equilibrium is difficult to establish and the gradient of $\beta$ continues to quickly change over most of the spatial domain. In larger domains, this effect persists about 10\% of the distance away from the emitting boundaries.}
\end{figure}

Phonon transport suffers from boundary scattering effects, and in systems which have strong non-equilibrium behavior, the boundaries can have a profound influence on solution behavior a finite distance into the geometric domain. In our method, $\beta$ has a strong gradient near any emitting boundary; both the equilibrium radiance and transport scalar flux are rapidly changing a finite distance away from a boundary. The magnitude of $\Phi^\textrm{T}$ near a boundary is a direct result of the boundary emission, where each outgoing ordinate of the angular intensity at an emissive boundary is assigned an incident phonon radiance (the adiabatic boundary condition). In turn, the magnitude of $\Phi^{00}$ depends on the solution of $\Phi^\textrm{T}$, but will not be equivalent near a boundary at short geometric domains due to the influence of an opposing boundary emission. To combat the strong boundary effects, we applied adaptive mesh refinement (AMR) using a gradient jump indicator. The gradient jump criteria takes the gradient norm of any specified solution quantity (in this case, we chose $\beta$) at the face between a mesh element and its' neighbor element. The elements are sorted by increasing error, and the elements which fall into a specified tolerance are refined. As we used triangular finite elements, a single level of refinement splits one element into four smaller elements of equivalent area, and so on; we specified a maximum of two refinement levels. All refinement took place a finite distance from the emissive boundaries, shown in Fig.~\ref{fig:AMR}, where the gradient of $\beta$ changes rapidly. The strongest boundary effects are observed in the 10 nm and 100 nm simulations, which is expected as phonons will nearly all be in the ballistic scattering regime at those domain sizes.

\begin{figure}
	\centering
	\includegraphics[scale=.20]{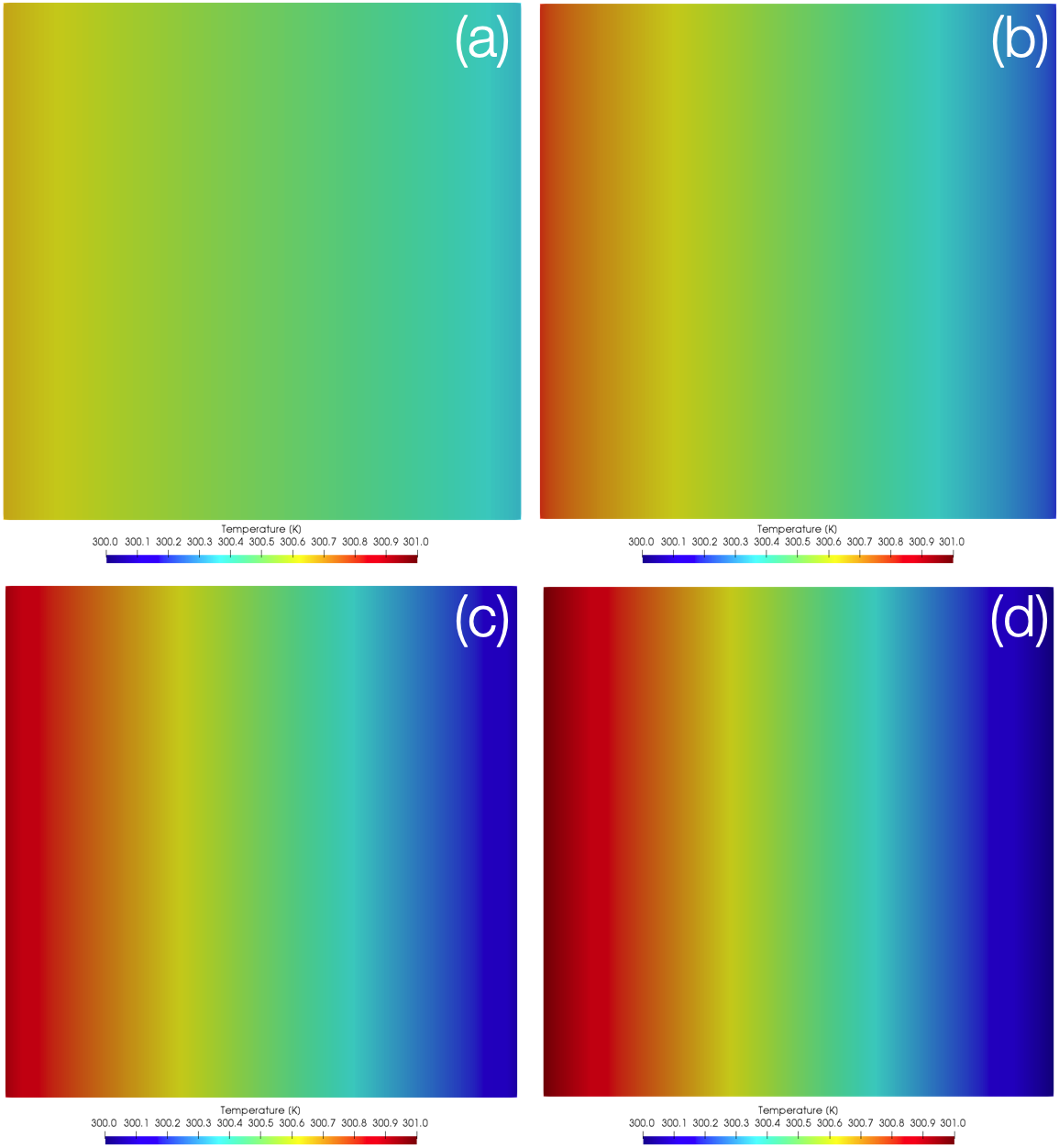}
\caption{\label{fig:quadTemp}The temperature distribution for four length scale (10 nm, 100 nm, 1 $\mu$m, 10 $\mu$m) cases in the $x$-$y$ plane; domain size increases from (a) - (d). In small domain sizes, equilibrium is difficult to establish. Phonon flux on the hot side is suppressed by phonons incident from the cold side; in turn this reduces temperature. All legend scales are equal.}
\end{figure}

The planar temperature distribution for four length scale (10 nm, 100 nm, 1 $\mu$m, 10 $\mu$m) cases are shown in Fig.~\ref{fig:quadTemp}; domain size increases from (a) - (d). In small domain sizes, equilibrium is difficult to establish. Phonon flux on the hot side is suppressed by phonons incident from the cold side, which reduces temperature. As the geometric domain increases in size, the Fourier limit is recovered and heat transport exists in an almost purely diffusive regime.

\begin{figure}
	\centering
	\includegraphics[scale=.20]{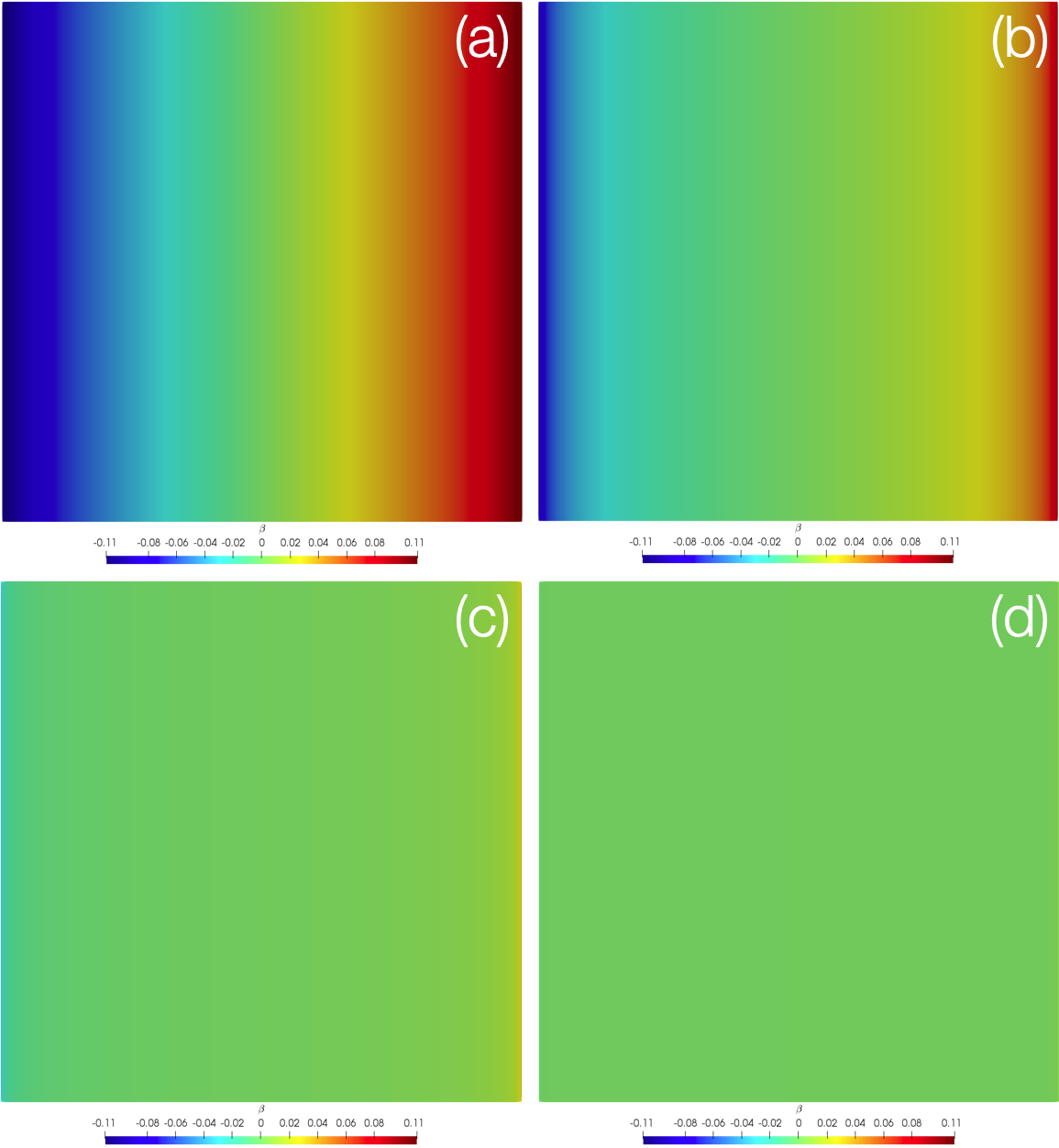}
\caption{\label{fig:betaMaps}$\beta$ distributions for four length scale (10 nm, 100 nm, 1 $\mu$m, 10 $\mu$m) cases in the $x$-$y$ plane; domain size increases from (a) - (d). A measure of local thermodynamic equilibrium, $\beta$ is nearly zero at large domain sizes, indicative of an established equilibrium. All legend scales are equal.}
\end{figure}

In Figure~\ref{fig:betaMaps} the distribution of $\beta$ is shown for various domain sizes. The relationship between a system unable to establish equilibrium (Fig.~\ref{fig:betaMaps}a) to one in an equilibrate state (Fig.~\ref{fig:betaMaps}d) is stark -- $\beta$ is nearly constant in large systems as the difference between $\Phi^{00}$ and $\Phi^\textrm{T}$ reduces. $\beta$ also shows the effect of the fixed boundary temperature and the slight curvature effect present in the temperature distribution. The magnitude and sign of $\beta$ also affect the convergence of the transport problem. Phonon transport simulations are classified as ``purely scatterering'' meaning the ratio of the mean free paths on the left and right hand sides of Eq.~(\ref{eq:snSAAF}) are unity, and these physics cause convergence to lag in acoustically thick problems. The presence of $\beta$ affects the balance of Eq.~(\ref{eq:snSAAF}), and in simulations where $\beta$ is not approximately constant in the bulk, the disparity between $\Phi^{00}$ and $\Phi^\textrm{T}$ is also large, in correlation with Fig.~\ref{fig:fluxAndRadiance}. This energy redistribution was required to close the transport system -- without its presence, we observed a non-conserved heat flux. The inclusion of $\beta$ projects the residual energy; the phonon groups are analogous to networks of siphons drawing from a central plenum, the energy bath.
\FloatBarrier

The ratio $\zeta_{p,\eta} = \frac{\mathcal{D}}{\Lambda_{p,\eta}}$ is the acoustic thickness of the domain ($\mathcal{D}$) with respect to mean free path per polarization and spectral group. The value of $\zeta_{p,\eta}$ changes in each group and polarization; certain group and mode polarizations are acoustically thick and affect the convergence of the numerical simulation. Table~\ref{tab:acousticProp} contains required iterations for each simulation and bounding values of $\zeta_{p,\eta}$ for the polarization and mode simulated. An increasing domain size requires more iterations to converge the solution, shown in Table~\ref{tab:acousticProp}. This is a consequence of the acoustic thickness; in the set $\bra{p,\Lambda_{\eta}}$, the minimum value of $\Lambda_{\eta}$ is the limiting factor. Due to the coupled nature of the transport simulation, the convergence performance of ballistic phonons is hindered by the diffuse phonons; in a decoupled simulation ballistic phonons converge much faster than diffuse phonons. The acoustic thickness in each polarization and mode are of primary interest in the numerical transport simulation. Depending on mode and polarization, $\Lambda_{p,\eta}$ can vary greatly due to its associated $v_{p,\eta}$ and $\tau_{p,\eta}$, affecting the acoustic thickness and causing significant degradation of numerical convergence, shown in Table~\ref{tab:acousticProp}. As optic phonons do not contribute significantly to thermal conductivity in these simulations, elimination of optic modes would reduce the overall amount of iterations; the simulation converges only as fast as its' slowest converging group. For the TA phonons, the implementation of an acceleration scheme such as diffusion synthetic acceleration (DSA) could reduce the required number of iterations.

One of the drawbacks of the coupled temperature method is the implicit dependency of all phonon channels; that is, if one or more of the channels is very diffuse, the simulations require many iterations to converge, if convergence is achieved at all. We have demonstrated convergence on domains ranging from 10 nm to 10 $\mu$m, but encountered significant decrease in efficiency when domain size increases to thousands of times the smallest value of $\Lambda_{\eta}$ (large $\zeta_{p,\eta}$). This is expected, as phonons in diffuse channels are highly scattering. We have observed similar, slowly convergent behavior in previous decoupled transport simulations where $\Lambda$ was small compared to the domain size~\cite{HarterANS2016,HarterJHT2018}. The distinction in a coupled implementation is that ballistic phonons are throttled by the presence of diffuse phonons, and convergence behavior is significantly affected. The involvement of a numerical acceleration scheme such as nonlinear diffusion acceleration (NDA) may improve convergence properties; with NDA, the solution of a lower order (with no angular variable) diffusion problem informs a source term in a higher order transport problem. This is a proven technique in neutron transport~\cite{rsm,LewisMiller,accelAdams,accelCMFD-FEM}, but may require significant revision for use in a phonon transport simulation and is planned in a future study.

Zhang~\emph{et al.} simulate a similar temperature coupling through informing each discrete phonon group by the overall temperature of the domain, and were able to simulate large spatial domains, up to 100 $\mu$m~\cite{zhangBTE}. However, they neglected optic phonons in their coupled simulations, and so convergence properties of their approach may be artificially positive as convergence of phonon transport problems is limited by the smallest value of $\Lambda$, e.g., the most diffuse phonon groups. Multiple researchers~\cite{FengRuanReview,WillaWang,Stock2011,Hsu2015} have cited the necessity in simulating optic phonons due to the acoustic-optic coupling effect found in various materials -- all phonons are coupled through the total energy of the domain. The inclusion of optic phonons effectively places a muzzle on the TA and LA phonons and prevents them from carrying their maximal potential of heat. While we study Si, as did Zhang \emph{et al.}, the development of this transport framework is desired to be general enough in its approach to be used in simulating materials which may have strong acousto-optic coupling effects.

We performed simulations isolating LA and TA phonons to gain insight into how much optic phonons contribute to $\kappa_\textrm{eff}$ depending on the size of the geometric domain. Table~\ref{tab:O_comparison} shows the difference in $\kappa_\textrm{eff}$ with and without optic phonon contribution. There is a nontrivial difference in $\kappa_\textrm{eff}$ when domain size is small, but this effect diminishes as the simulation becomes more diffuse with increasing acoustic thickness. It is important to recognize the contribution of optic phonons in smaller spatial domains; at these sizes, optic phonons can exist ``agnostically'' between the ballistic and diffuse regimes. The presence of optic phonons causes significant degradation in the performance of the solver when comparing iterations between Table~\ref{tab:acousticProp} and \ref{tab:acousticProp_TL}. While optic phonons may not be required in large simulation cells due to overshadowing influence of LA and TA phonons, it is nonetheless important to consider their contribution $\kappa_\textrm{eff}$, and subsequent effect on solver performance.

\begin{table}[!htbp]
\caption{\label{tab:O_comparison} Difference in $\kappa_{\textrm{eff}}$ with and without optic phonons for $S_{8,F}$ simulation. Thermal conductivity is in units of $\sq{\textrm{W}\cdot\textrm{m}^{-1}\cdot\textrm{K}^{-1}}$.}
\begin{ruledtabular}
\begin{tabular}{lcdr}
$L$ [nm] &
{$\kappa_{\textrm{LA,TA,O}}$} &
\multicolumn{1}{c}{$\kappa_{\textrm{LA,TA}}$} &
$\Delta \kappa\rnd{\%}$ \\
\colrule
10    & 6.1   & 5.2   & 14.8  \\
100   & 30.3  & 28.2  & 6.9 \\
1000  & 90.8  & 88.4  & 2.6  \\
10000 & 168.6 & 166.7 & 1.1  \\
\end{tabular}
\end{ruledtabular}
\end{table}
\begin{table}[!htbp]
\caption{\label{tab:acousticProp} $S_{8,F}$ with LA, TA, O phonons.}
\begin{ruledtabular}
\begin{tabular}{lcdr}
$L$ [nm] &
GMRES Iterations &
\multicolumn{1}{c}{$\min\rnd{\zeta_{p,\eta}}$} &
{$\max\rnd{\zeta_{p,\eta}}$} \\
\colrule
10    & 29    & 0.0032 & 3.175  \\
100   & 59    & 0.032  & 31.75 \\
1000  & 1276  & 0.32   & 317.5  \\
10000 & 46600 & 3.2    & 3175  \\
\end{tabular}
\end{ruledtabular}
\end{table}
\begin{table}[!htbp]
\caption{\label{tab:acousticProp_TL} $S_{8,F}$ with LA and TA phonons.}
\begin{ruledtabular}
\begin{tabular}{lcdr}
$L$ [nm] &
GMRES Iterations &
\multicolumn{1}{c}{$\min\rnd{\zeta_{p,\eta}}$} &
{$\max\rnd{\zeta_{p,\eta}}$} \\
\colrule
10    & 28   & 0.0032 & 0.901  \\
100   & 41   & 0.032  & 9.01 \\
1000  & 114  & 0.32   & 90.1  \\
10000 & 3306 & 3.2    & 901  \\
\end{tabular}
\end{ruledtabular}
\end{table}
\section{Conclusions}
We have developed a temperature coupled, spectral method for simulating effective thermal conductivity from nano- to micro-scale using material properties for homogeneous silicon obtained from \emph{ab initio} density functional theory. In this study we simulated 2D geometric domains to characterize algorithmic and transport performance. The extension of this method to simulate three-dimensional structures is trivial, but will demand additional spatial and angular degrees of freedom. Angular degrees of freedom scale as $N(N+2)$ for 3D and $N(N+2)/2$ for 2D, where $N$ is the chosen order of the discrete ordinates angular discretization. Spatial degrees of freedom scale with the nodal count in the selected finite element type, e.g., 3 nodes in 2D for a triangular element, 4 nodes in 3D for a tetragonal element. Computational cost will scale proportional to degrees of freedom in the equation system. We demonstrated correlation between the geometric dependency of heat flux and thermal conductivity, and our results for $\kappa_\textrm{eff}$ in Si are in good agreement with those available in the open literature. Convergence properties are negatively impacted by the presence of optic phonons; the diffuse characteristics of optic phonons increase the acoustic thickness of simulations. A closure term, $\beta$, was developed to conserve the local system energy, and this fraction is projected as a source term in each phonon group, providing additional systemic coupling through the energy bath. The range and sign of $\beta$ indicates the degree of disorder between the transport system and the total energy of the phonon bath. In small geometric domains, the importance of using BTE simulations to model heat transport cannot be overstated; the difference between $\Phi^\textrm{T}$ and $\Phi^{00}$ is large where equilibrium cannot be established, and spatial effects dramatically change temperature distributions and heat flux. Previous works have shown optic phonons to contribute little to overall heat flux in silicon~\cite{detMajumdar1,detMajumdar2,zhangBTE,detYBM1}. However, geometric domains less than $100$ nm in size (Table \ref{tab:O_comparison}), and in devices where precision is critical, optic phonons should be included in simulation to accurately compute $\kappa_{\textrm{eff}}$. This work focused on purely homogeneous materials to characterize a new method, but our approach can be extended to bulk systems with voids (porous materials for thermoelectric devices) and heterogeneous systems (systems with inclusions, nuclear fuel with fission product defects) quite readily with the addition of an interface physics model. Subsequent improvement on this work would include simulation of transference between polarization branches, and the additional modeling of 3-phonon scattering through the relaxation time parameter included in the total relaxation time via Matthiessen's rule. We are currently developing diffusion acceleration techniques adapted from neutron transport methods in order to speed up convergence properties as domains become extremely acoustically thick.

Detailed thermal transport requires development of high precision tools with which to perform modeling and simulation efforts. In turn, these efforts can guide the design of nano- and micro-scale devices, and the prediction of thermal behavior at the engineering scale. Perturbations at the atomic-molecular scale drive changes in the microstructure; \emph{ab initio} driven BTE simulations are a promising development in a scale-bridging framework, representing one element in the larger problem of informing engineering scale simulations.
\begin{acknowledgments}
The authors thank Sebastian Schunert and Yaqi Wang at INL for their invaluable knowledge and assistance. This research made use of the resources of the High-Performance Computing Center at Idaho National Laboratory, which is supported by the Office of Nuclear Energy of the U.S. Department of Energy and the Nuclear Science User Facilities under Contract No. DE-AC07-05ID14517. We also acknowledge the support of Pacific Northwest National Laboratory.
\end{acknowledgments}
\bibliography{masterBiblio}
\end{document}